%% file: paper_clean.tex
\documentclass[%
 reprint,
 amsmath,amssymb,
aps,
prb,
]{revtex4-2}

\usepackage{graphicx}
\usepackage{dcolumn}
\usepackage{bm}
\usepackage{xcolor}

\begin{document}

\preprint{APS/123-QED}

\title{High temperature and pressure thermoelasticity of hcp metals from ab initio quasi-harmonic free energy calculations: the  beryllium case}

\author{Xuejun Gong}
 \affiliation{International School for Advanced Studies (SISSA), Via Bonomea 265, 34136, Trieste, Italy}
 \affiliation{IOM - CNR, Via Bonomea 265, 34136, Trieste, Italy}

\author{Andrea Dal Corso}
 \affiliation{International School for Advanced Studies (SISSA), Via Bonomea 265, 34136, Trieste, Italy}
 \affiliation{IOM - CNR, Via Bonomea 265, 34136, Trieste, Italy}

\date{\today}

\begin{abstract}
We present a systematic ab initio study of the temperature and pressure dependent thermoelastic properties of hcp beryllium within the quasi-harmonic approximation (QHA). The accuracy of the Zero Static Internal Stress Approximation (ZSISA) and of the volume-constrained ZSISA that are widely applied in ab initio thermodynamic calculations are quantified. Particularly, the effect of ZSISA for the calculation of $C_{11}$ and $C_{12}$ is compared with a novel numerical approach which minimizes the free energy with respect to the atomic positions at each strain.  In beryllium, minor deviations are found within ZSISA, which gives ECs in good agreement with the full free energy minimization (FFEM).
A substantial difference is found between QHA and the quasi-static approximation (QSA), with the former closer to experiments.
Within QSA, we compare the ECs obtained by interpolating from a set of geometries along the ``stress-pressure'' isotherm at $0$ K (within V-ZSISA) with a more general interpolation on a two-dimensional grid of crystal parameters which allows the calculation of the ECs along the $0$ kbar isobar. This paper provides a practical approach for the investigation of the thermoelastic properties of hcp metals at extreme conditions.
 \end{abstract}

\maketitle


\section{Introduction}
Beryllium is a lightweight metal with a very low density, high elasticity and thermal conductivity, extremely low Poisson ratio and several other noteworth physical properties that make it quite attractive for applications in aircrafts, satellites, and spacecraft. It is also used in nuclear power industry as a neutron reflector and moderator.
Its thermodynamic properties are well explored, experimentally~\cite{arblaster_thermodynamic_2016,bodryakov_correlation_2014,stedman_phonon_1976,lazicki_high-pressure--temperature_2012} and theoretically~\cite{luo_ab_2012,hao_first-principle_2012,kadas_temperature-dependent_2007,sinko_relative_2005,robertMultiphaseEquationState2010,shao_temperature_2012,wu_high-pressure_2021}, but the knowledge of its elastic constants (ECs) is still improvable. 

Room temperature
ECs, measured several times (see Ref.~\cite{migliori_berylliums_2004} for a recent account) have been calculated at $0$ K by many authors. As one of us discussed previously~\cite{dal_corso_elastic_2016}, the reported results are not always in agreement among themselves, sometimes due to different numerical techniques but sometimes also due to the different treatment of internal relaxations.

Pressure dependent ECs at $0$ K have been calculated in Refs.~\cite{sinko_relative_2005,hao_first-principle_2012,luo_ab_2012} with the first two papers in substantial agreement while the third that predicts a somewhat different pressure dependence.

For the temperature dependent elastic constants (TDECs), two sets of experimental data exist at room pressure. The first~\cite{smith_elastic_1960} covering the low temperature range from $0$ K to $300$ K and the second~\cite{rowlands_determination_1972} the range from $298$ K to $573$ K. Ref.~\cite{rowlands_determination_1972} reported a quite strong decrease in ECs with temperature, a fact that motivated further theoretical investigations~\cite{kadas_temperature-dependent_2007,robertMultiphaseEquationState2010,shao_temperature_2012} using the quasi-static approximation (QSA) in Ref.~\cite{kadas_temperature-dependent_2007} and the quasi-harmonic approximation (QHA) in Refs.~\cite{robertMultiphaseEquationState2010,shao_temperature_2012}. None of these studies could obtain the rapid decrease of the ECs claimed by Ref.~\cite{rowlands_determination_1972}
and a reexamination of the experimental data was suggested. Ref.~\cite{nadalElasticModuliBeryllium2010} measured the compressional and shear sound velocities of polycrystalline beryllium and derived the bulk and shear modulus from them. Although the experimental errors are still quite large, the results are more in line with the theoretical data than with Ref.~\cite{rowlands_determination_1972}.

At high pressure and high temperature the situation is even more obscure. We are not aware of any experimental or theoretical paper available so far.

In this paper we reexamine the TDECs of beryllium focusing on the analysis of the effects of the common approximations made for studying the ECs of anisotropic solids: the zero static internal stress  approximation (ZSISA)~\cite{allanZeroStaticInternal1996a} and the constant volume 
(V-ZSISA)~\cite{masukiFullOptimizationQuasiharmonic2023a} approximation (also called the statically constrained quasi-harmonic approximation~\cite{carrierFirstprinciplesPredictionCrystal2007}). Within ZSISA one avoids the calculation of the free energy as a function of the atomic positions in strains that decrease the symmetry enough to let the atoms free to move. For each strain, the atomic positions are calculated at $0$ K from energy minimization and the free energy is computed at one atomic configuration. Using the V-ZSISA the equilibrium configurations are obtained at $0$ K by optimizing (using energy) the crystal parameters in a set of volumes $V_i$ (or pressure $p_i$) and computing the free energy only on the optimized geometries. 

After the optimization of the crystal parameters and atomic positions,
the ECs can be calculated within the QSA (from the second strain derivatives of the energy) or within the QHA (from the second strain derivatives of the free energy).
We report both the QSA and QHA TDECs calculated within V-ZSISA along the ``stress-pressure'' $0$ K isotherm determined so that the stress is a uniform pressure along it. 

The effect of V-ZSISA is tested on the QSA TDECs
by identifying in the plane of parameters $a$ and $c/a$ the isotherm at $1500$ K and
interpolating the ECs along the ``stress-pressure'' isotherm at $0$ K (within V-ZSISA) or along the correct isobar at $0$ kbar that joins the two isotherms.

In hexagonal close packed (hcp) crystals, relaxation of atomic positions affects only the ECs $C_{11}$ and $C_{12}$. On these, we test the ZSISA, by comparing its predictions with the ECs calculated with atomic positions that minimize the free energy.

We find that both V-ZSISA and ZSISA in beryllium are accurate and have only minor effects on the final QHA
ECs.

As in other metals~\cite{malica_quasi-harmonic_2020,malica_quasi-harmonic_2021,gong_pressure_2024,gong_ab_2024}, QHA gives results closer to experiment than the QSA, but even if the QHA gives a faster decrease with temperature of $C_{11}$, $C_{33}$, and $C_{44}$, the derivatives with respect to temperature of these ECs are still lower than in experiment and in substantial agreement with previous calculations.

Finally, we present the pressure-dependent QHA ECs at $4$ K, $500$ K, and at $1000$ K, in hopes that these theoretical data can help and stimulate the experimental measurement of these quantities.

\begin{figure}
\centering
\includegraphics[width=\linewidth]{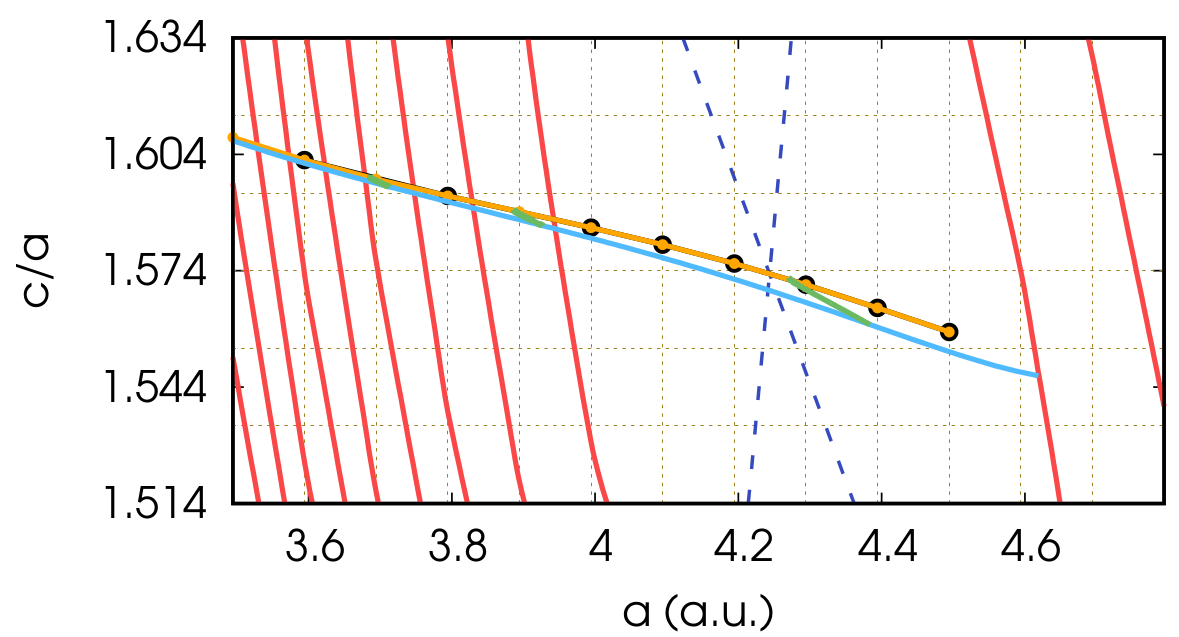}
\caption{Contours of constant total energy (red lines) plotted in the plane $a$ and $c/a$. The two blue dashed straight lines intersect at the position of the energy minimum. The orange curve is the ``stress-pressure'' isotherm at $0$ K. The light-blue curve is the ``stress-pressure'' isotherm at $1500$ K.
The three green lines show the isobars
at $0$ kbar, $500$ kbar, and $1000$ kbar for temperatures going from $0$ K to $1500$ K. 
Points on the orange curve shows the values of $a$ and $c/a$ in which we have computed the quasi-harmonic TDECs. The $0$ K ECs as well as the phonon dispersions have been calculated in these points and also in all the points of the two dimensional grid shown with dotted lines.}
\label{fig:energy}
\end{figure}

\section{Theory}
\subsection{Thermodynamics and elastic constants}

In this paper, the \texttt{thermo\_pw}~\cite{dal_corso_thermo_pw_2022} software developed by ourselves is employed to calculate all thermodynamic properties.
The QHA, as implemented in \texttt{thermo\_pw}, has been  discussed in previous publications~\cite{dal_corso_elastic_2016,palumboLatticeDynamicsThermophysical2017c,pal2,malica_temperature_dependent_2019,malica_quasi-harmonic_2020,malica_temperature_2020,malica_quasi-harmonic_2021}. Here, we summarize the main formulas and discuss the thermodynamic relationships
needed for computing the ECs of hcp solids. Except for a few relationships that are more easily written in cartesian coordinates, we will use the
Voigt notation with indices going from $1$
to $6$.

Within QHA, the Helmholtz free energy $F(\xi,T)$ of a solid is a function
of temperature $T$ and (unit cell) parameters $\xi$
that in the hexagonal lattice are $a$ and $c/a$. It can be written as the sum
of three contributions:
\begin{equation}
F(\xi,T)= U(\xi) + F_{ph}(\xi,T) + F_{el}(\xi,T),
\label{eq:free_ener}
\end{equation}
where $U(\xi)$ is the static energy, 
$F_{ph}(\xi,T)$ is the vibrational free energy, and
$F_{el}(\xi,T)$ is the electronic excitations contribution to the free energy.
$U(\xi)$ is computed via density functional theory (DFT), $F_{ph}(\xi,T)$ is
written in terms of the phonon frequencies
${\omega}_{\eta}({\bf q},\xi)$:
\begin{eqnarray} \label{equ4}
F_{vib}(\xi, T) & =& \frac{1}{2N} \sum_{\mathbf q \eta} \hbar \omega_{\eta} \left(\mathbf q,
\xi \right) \nonumber\\
& +& {\frac{1}{N \beta}} \sum_{\mathbf q \eta} \ln \left[1 - \exp \left(- \beta \hbar \omega_{\eta}(\mathbf q, \xi)\right) \right],
\label{eq:fvib}
\end{eqnarray}
and $F_{el}(\xi,T)$ can be computed within the rigid bands approximation from the 
electronic density of states (see
Ref.~\cite{malica_quasi-harmonic_2021}). In beryllium we
expect small effects of electronic excitations~\cite{kadas_temperature-dependent_2007} and
in this paper we do not consider them.
In Eq.~\ref{eq:fvib}, $\hbar$ is the reduced Planck's constant,
$\beta={\frac{1}{k_B T}}$, where $k_B$ is the Boltzmann constant, ${\bf q}$ are the phonon wavevectors and $\eta$ indicates the different modes. $N$ is the number of cells of the solid (equal also to the number of phonon wavevectors {\bf q}). These free energies are computed for a grid of parameters $\xi_i=(a_i,c_i/a_i)$, $i=1,N_p$. $U(\xi)$ as well as the vibrational free energy are interpolated by a fourth-degree polynomial.

Considering the stress tensor $\boldsymbol{\sigma}$ as a fixed set of parameters, and the strain as a function of the crystal parameters minimization of the functional:
\begin{equation}
G_{\boldsymbol{\sigma}}(\xi,T)=F(\xi,T)-V\sum_j \boldsymbol{\sigma}_j \epsilon_j
\label{gibbs}
\end{equation}
with respect to the parameters $\xi$ gives the EOS:
\begin{equation}
\boldsymbol{\sigma}_j={1\over V} {\frac{\partial F(\xi,T)}{\partial \epsilon_j}}.
\label{eq:p}
\end{equation}
Hence the crystal parameters that minimizes $G_{\boldsymbol{\sigma}}(\xi,T)$ are those that give stress $\boldsymbol{\sigma}$. Using for the stress a uniform pressure we find the crystal parameters at any pressure  and temperature ($\xi_p(T)$). From the $\xi_p(T)$ we can compute also the volume as a function of $p$ that is the equation of state (EOS): $V(p,T)=V(\xi_p(T))$.

Using $V(p,T)$ we obtain the volume thermal expansion $\beta(p,T)$
at pressure $p$ as:
\begin{equation}
\beta(p,T)= {\frac{1}{V(p,T)}} {\frac{\partial V(p,T)}{\partial T}} \Bigg |_p.
\label{eq:beta}
\end{equation}
For an hexagonal system, the thermal expansion tensor is diagonal and has two different components.
We get:
\begin{eqnarray}
\alpha_{1} &=& \alpha_{2} = {1\over a} {d a \over d T}, \\
\alpha_{3} &=& {1\over c} {d c \over d T}. 
\end{eqnarray}

The isothermal ECs are calculated from the second strain
derivatives of the free energy.
\begin{equation}
\tilde C^T_{ij}= {1\over V} { \partial^2 F \over \partial \varepsilon_{i}
\partial \varepsilon_{j}} \Bigg |_T,
\label{tdec}
\end{equation}
Actually using the following
five strain types: $(\epsilon,0,0,0,0,0)$,
$(0,0,\epsilon,0,0,0)$, $(\epsilon,0,\epsilon,0,0,0)$,
$(\epsilon,\epsilon,0,0,0,0)$, and $(0,0,0,\epsilon,0,0)$, 
${1\over V} {\partial^2 F \over \partial \epsilon^2}$
is equal to 
$\tilde C_{11}$, $\tilde C_{33}$, $\tilde C_{11} +\tilde C_{33} + 2 \tilde C_{13}$, $2 \tilde C_{11} + 2 \tilde C_{12}$, and $\tilde C_{44}$ respectively.
When the equilibrium reference configuration has a non vanishing stress $\sigma^{(0)}_{i}$ (or $\sigma^{(0)}_{ij}$ in cartesian notation), the stress-strain ECs $C^T_{ij}$ are obtained as (in cartesian notation)~\cite{barron_second-order_1965}:
\begin{eqnarray}
C^T_{ijkl} =  \tilde C^T_{ijkl} &-& 
{1\over 2} \Big(2 \sigma^{(0)}_{ij} \delta_{kl} 
-{1\over 2} \sigma^{(0)}_{ik} \delta_{jl} 
-{1\over 2} \sigma^{(0)}_{il} \delta_{jk} \nonumber \\
&-&{1\over 2} \sigma^{(0)}_{jk} \delta_{il} 
-{1\over 2} \sigma^{(0)}_{jl} \delta_{ik} \Big).
\label{eqsd2}
\end{eqnarray}
An hexagonal lattice with an arbitrary $a$ and $c/a$ has
a diagonal stress tensor with two equal components $\sigma^{(0)}_{1}=\sigma^{(0)}_{2}$, while $\sigma^{(0)}_{3}$ can be different.
From Eq.~\ref{eqsd2} we find
$C^T_{11}=\tilde C^T_{11}$, $C^T_{33}=\tilde C^T_{33}$
while $C^T_{12}=\tilde C^T_{12}-\sigma^{(0)}_1$, $C^T_{21}=\tilde C^T_{21}-\sigma^{(0)}_1$,
$C^T_{13}=\tilde C^T_{13}-\sigma^{(0)}_1$,
$C^T_{31}=\tilde C^T_{31}-\sigma^{(0)}_3$,
$C^T_{44}=\tilde C^T_{44}+{1\over 4} (\sigma^{(0)}_1+\sigma^{(0)}_3)$. Since $\tilde C_{ij}$ is symmetric
in the exchange of the two indices, $C^T_{ij}$ is not.
For an hexagonal lattice we have $C^T_{12}=C^T_{21}$,
but $C^T_{31}\ne C^T_{13}$. Symmetry is recovered only along the ``stress-pressure'' isotherm
where $\sigma^{(0)}_1=\sigma^{(0)}_3=-p$.
Along this curve Eq.~\ref{eqsd2} becomes
(in Cartesian notation):
\begin{equation}
C^T_{ijkl} = \tilde C^T_{ijkl} + {p \over 2} \left(2 \delta_{i,j} \delta_{k,l}
- \delta_{i,l} \delta_{j,k} - \delta_{i,k} \delta_{j,l}  \right).
\label{eq:correct_p}
\end{equation}
The second derivatives of the free energy are calculated as described in 
Ref.~\cite{dal_corso_elastic_2016} taking as equilibrium configuration a subset of parameters $\xi_i$ along the ``stress-pressure'' $0$ K isotherm. The values of $\xi_i$ along this curve are given in the supplementary material, together with the pressure present in each configuration.
The ECs at any other set of parameters $\xi_p$ at temperature $T$ and pressure $p$ are obtained by projection on the `stress-pressure'' $0$ K isotherm ($a(T)$ is unchanged while $c/a(T)$ is substituted with $c/a(a(T))$) and interpolation by a fourth-degree polynomial.

Adiabatic ECs are calculated from the isothermal ones as:
\begin{equation}
C^S_{ij}=C^T_{ij} + {T V b_{i} b_{j} \over C_V},
\end{equation}
where $b_{i}$ are the thermal stresses:
\begin{equation}
b_{i} = - \sum_{j} C^T_{ij} \alpha_{j}.
\end{equation}

\begin{figure}
\centering
\includegraphics[width=\linewidth]{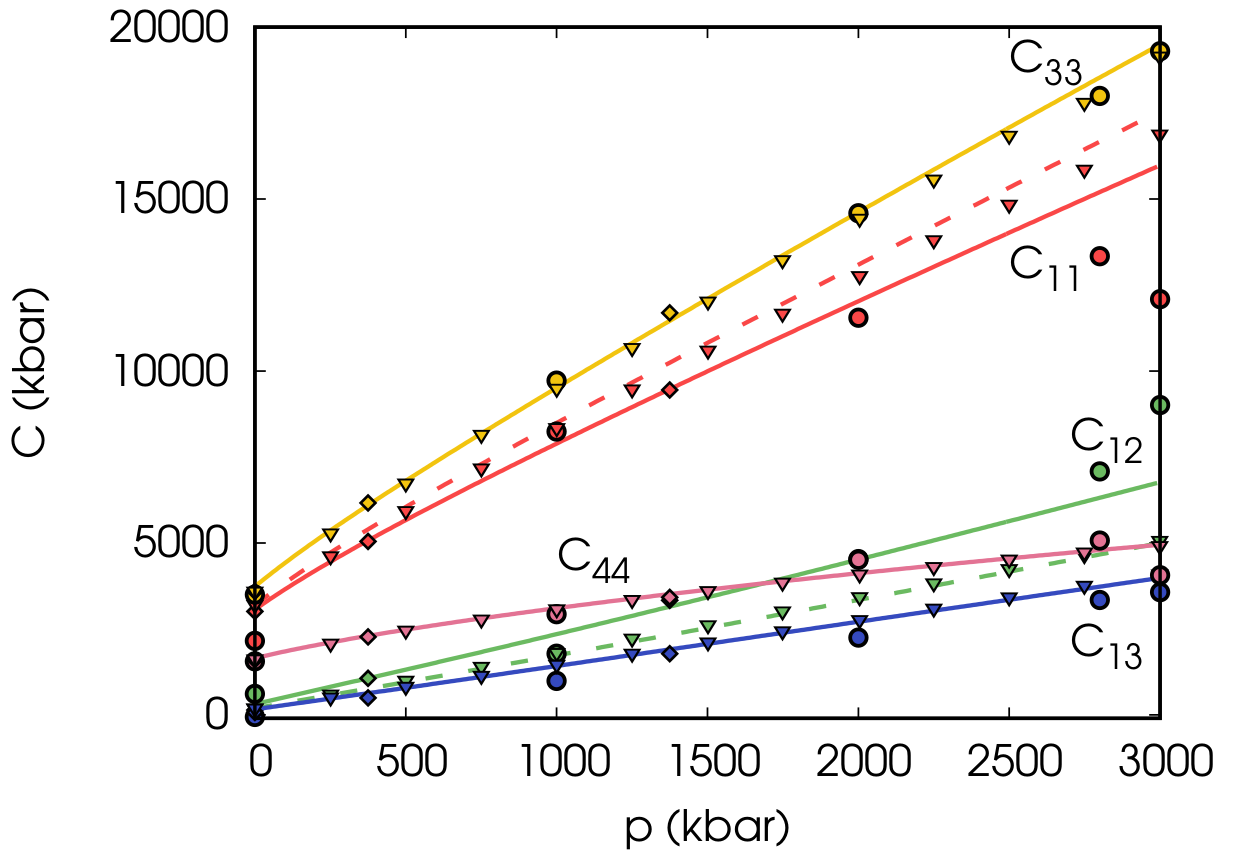}
\caption{Stress-strain elastic constants of Be as a function of pressure at $0$ K (continous lines) compared with previous calculations of Ref.~\cite{sinko_relative_2005} (diamonds), Ref.~\cite{hao_first-principle_2012} (circles), and Ref.~\cite{luo_ab_2012} (triangles). The dashed lines show $C_{11}$ and $C_{12}$ obtained by keeping the ions fixed at the uniformly strained positions.}
\label{fig:elastic_p}
\end{figure}
\subsection{HCP internal relaxations}
The application of a strain $(\epsilon,0,0,0,0,0)$
to the hcp structure transforms the hexagonal lattice into a base centered orthorhombic lattice and the positions of the two atoms in the unit cell are no more constrained in the $y$ direction.
The energy can be written in the form
\begin{equation}
E(\epsilon, y) = {1\over 2} V C_{11}^{(0)} \epsilon^2
+ \Lambda \epsilon y + {1\over 2} \mu \omega^2 y^2 + E(0,0),
\label{energy1}
\end{equation}
where $C_{11}^{(0)}$ is the frozen ion ECs obtained by keeping the two atoms of the hcp unit cell in the strained position. In this
equation $y$ is the deviation of the positions of the two atoms from their strained position
(that for the $y$ coordinate coincides with the equilibrium position)
$a{1\over 2 \sqrt{3}}$ where $a$ is the hexagonal unstrained lattice parameter (We refer to Fig.2 of Ref.~\cite{dal_corso_elastic_2016} for an illustration of the geometry).
By minimizing the energy with respect to $y$ we find:
\begin{equation}
    y= - {\Lambda \epsilon \over \mu \omega^2},
\label{yzsisa}    
\end{equation}
that inserted in Eq.~\ref{energy1} gives the correction to the $C^{(0)}_{11}$ EC.

We find:
\begin{equation}
    C_{11}= C^{(0)}_{11} - {\Lambda^2 \over V \mu              \omega^2}.
\end{equation}

Similarly, within the QHA approximation, 
we can use the free energy instead of the
energy and write:
\begin{eqnarray}
F(\epsilon, y, T) &=& {1\over 2} V C_{11}^{F (0)} (T) \epsilon^2
+ \Lambda^F(T) \epsilon y + {1\over 2} \mu \omega_F^2(T) y^2 \nonumber \\ &+& F(0,0,T),
\label{fhcps1}
\end{eqnarray}
By minimizing the free energy at each 
temperature we find:
\begin{equation}
    y^F= - {\Lambda^F(T) \epsilon \over \mu \omega^2_F(T)},
\label{ffem}    
\end{equation}
and we obtain the correction to the $C^{F (0)}_{11}$
EC:
\begin{equation}
    C^F_{11}(T)= C^{F (0)}_{11}(T) - {\Lambda^{F 2}(T) \over V \mu \omega_F^2(T)}.
\label{dc11hcp}
\end{equation}
Using for $y$ Eq.~\ref{yzsisa} instead of Eq.~\ref{ffem} is the ZSISA approximation.

\subsection{Elastic constants computation beyond ZSISA}
The equations in the previous subsection provide a method to compute the ECs accounting for internal relaxations without ZSISA. Similarly to what was done in Ref.~\cite{malicaFinitetemperatureAtomicRelaxations2022}, for each strain, it is possible to calculate the free energy for a finite number of atomic positions. The free energy as a function of strain and atomic coordinates is then interpolated at each temperature with a polynomial as in Eq.~\ref{fhcps1}. The mixed second derivatives $\Lambda^F(T)$ and the frequencies  $\mu \omega_F^2(T)$
are calculated from the interpolating polynomial and the correction to the frozen ions ECs derived from Eq.~\ref{dc11hcp}. 

In this paper, we propose an alternative method to compute the ECs in presence of internal relaxation that we call full free energy minimization (FFEM). For each strain, the energy (or free energy) as a function of the internal position $y$ is interpolated with a second or fourth degree polynomial and the minimum is found. The value of the minimum (free-) energy is assigned to the given strain and used to calculate the TDECs via Eq.~\ref{tdec}.
This approach, which at $0$ K is equivalent to the relaxed-ions calculation, has the advantage that it can be carried out at any temperature and, at variance with the approach of Ref.~\cite{malicaFinitetemperatureAtomicRelaxations2022}, does not require the knowledge of the form of the interpolating polynomial, that might be structure dependent and has to be analyzed on a case by case basis. Therefore, using the full free energy minimization (FFEM) we obtain the relaxations and ECs beyond the ZSISA and compare them with the ZSISA ones.  A similar method that goes beyond ZSISA has been applied for the calculation of the internal thermal expansion of ZnO~\cite{liuInternalExternalThermal2018}.

\section{Technical details}
The calculations presented in this work are done by using DFT as implemented in the Quantum ESPRESSO (QE) package.~\cite{qe1, qe2} The exchange and correlation functional is the LDA.~\cite{lda}
We employ a plane-wave basis with the pseudopotential \texttt{Be.pz-n-vbc.UPF} obtained from the QE website. This pseudopotential has the $2s$ states in valence, while the $1s$ electrons are frozen in the core and accounted for by the nonlinear core correction~\cite{louie_nonlinear_1982}.
For the wave functions and charge density cutoffs, we use $35$ Ry and $140$ Ry respectively. The Fermi surface has been dealt with by the smearing approach of Methfessel and Paxton~\cite{mp} with a smearing parameter $\sigma = 0.02$ Ry. With this smearing, the Brillouin zone integrals give reasonable values of the ECs with a $64 \times 64 \times 40$ {\bf k}-point mesh.

We first determine the ``stress-pressure'' $0$ K isotherm in the crystal parameters space by computing the total energy in a mesh of $14 \times 7$ grid of values of $a$ and $c/a$ covering a pressure range from about $-200$ kbar to $1800$ kbar. On this grid of geometries, we compute also the phonon dispersions and the $0$ K ECs. This give us
the thermal expansion tensor and the
``stress-pressure'' isotherm at any temperature, as well as the QSA ECs without the V-ZSISA approximation.

Along the ``stress-pressure'' isotherm at $0$ K, we choose $11$ values of $a$ and $c/a$ as given in Tab. I in the supplementary material. In these geometries we compute the phonon dispersions, the free energy and the $0$ K ECs. In $8$ of these $11$ geometries we also compute the QHA TDECs as second strain derivatives of the free energy. These ECs are then used to interpolate the ECs for any other pressure and temperature within the V-ZSISA approximation. The $8$ reference geometries have $i=2$, $4$, $6$, $7$, $8$, $9$, $10$, and $11$ (where geometry $1$ is the point at highest pressure) and the QHA TDECs are calculated by $5$ strain types that lead to base centered orthorhombic (strain types $1$ and $3$), hexagonal (strain types $2$ and $4$) and monoclinic (strain type $5$) lattices. Each strain type is sampled by 6 strains, from $\epsilon=-0.0125$ to $\epsilon=0.0125$ with a stepsize $\delta \epsilon= 0.005$. Each of the $30\times 8=240$ strained configurations requires calculations of the phonon frequencies by density functional perturbation theory (DFPT)~\cite{rmp, dfptPAW} to obtain the dynamical matrices on a $6 \times 6 \times 6$ \textbf{q}-point grid. This grid 
leads to $28$ inequivalent {\bf q}-points in the
hexagonal cell, $52$ in the
base centered orthorombic cell and to 
$68$ in the monoclinic cell.

To calculate FFEM ECs, for each equilibrium geometry, free energies are needed on $78$ strained configurations.
This number is determined by considering that for strain type $1$ and $3$, six values
of strain $\epsilon$ are sampled and, in addition, we calculate $5$ different values of $y$.  Therefore, the five strain types of hcp structure will require $30+6+30+6+6=78$ phonon dispersions.

The dynamical matrices calculated by DFPT are Fourier interpolated into a $200 \times 200 \times 200$ \textbf{q}-point mesh to evaluate the free energy and
its strain derivatives.
The calculations are all performed on the Leonardo supercomputer at CINECA with a GPU version of \texttt{thermo\_pw} that optimizes some routines of QE for problems with dense {\bf k}-points sampling in metallic systems~\cite{gong_dalcorso_opt}.
Please refer to the supplementary material for a workflow of the present 
calculations.~\cite{supplemental}

Recently, some methods to calculate the dynamical matrices in strained configurations~\cite{masukiFullOptimizationQuasiharmonic2023a} or to reduce the number of calculated phonon dispersions needed for QHA thermal expansion~\cite{rostamiApproximationsFirstprinciplesVolumetric2024} and for QHA TDECs~\cite{ mathisGeneralizedQuasiharmonicApproximation2022} have been proposed. It could also be useful to try them in order to speed up the calculations in our problem.

\section{Results and Discussion}

\begin{table*}
  \input{table2.tex}
\end{table*}

\begin{figure}
\centering
\includegraphics[width=\linewidth]{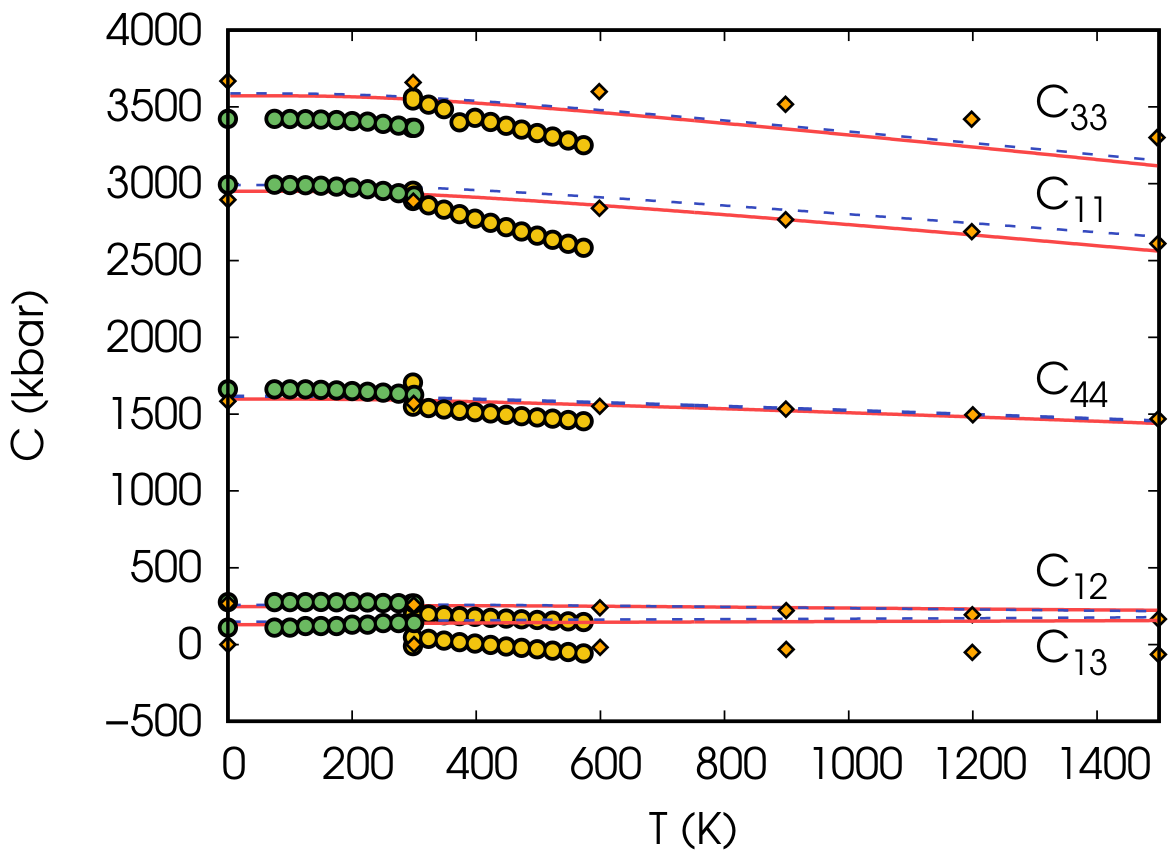}
\caption{Adiabatic LDA elastic constants of Be as a function of temperature calculated within the QSA (red lines) along the $0$ kbar isobar (with ZSISA atomic positions). For comparison we have reported also the QSA elastic constants interpolated (within V-ZSISA) only on the ``stress-pressure'' isotherm at $0$ K (dashed blue lines). The dots are the experimental points of Ref.~\cite{rowlands_determination_1972} (yellow dots) and \cite{smith_elastic_1960} (green dots). Diamond are the theoretical PBE QSA calculation of Ref.~\cite{kadas_temperature-dependent_2007}.}
\label{fig:elastic_qsa_all_1d}
\end{figure}

\begin{figure}
\centering
\includegraphics[width=\linewidth]{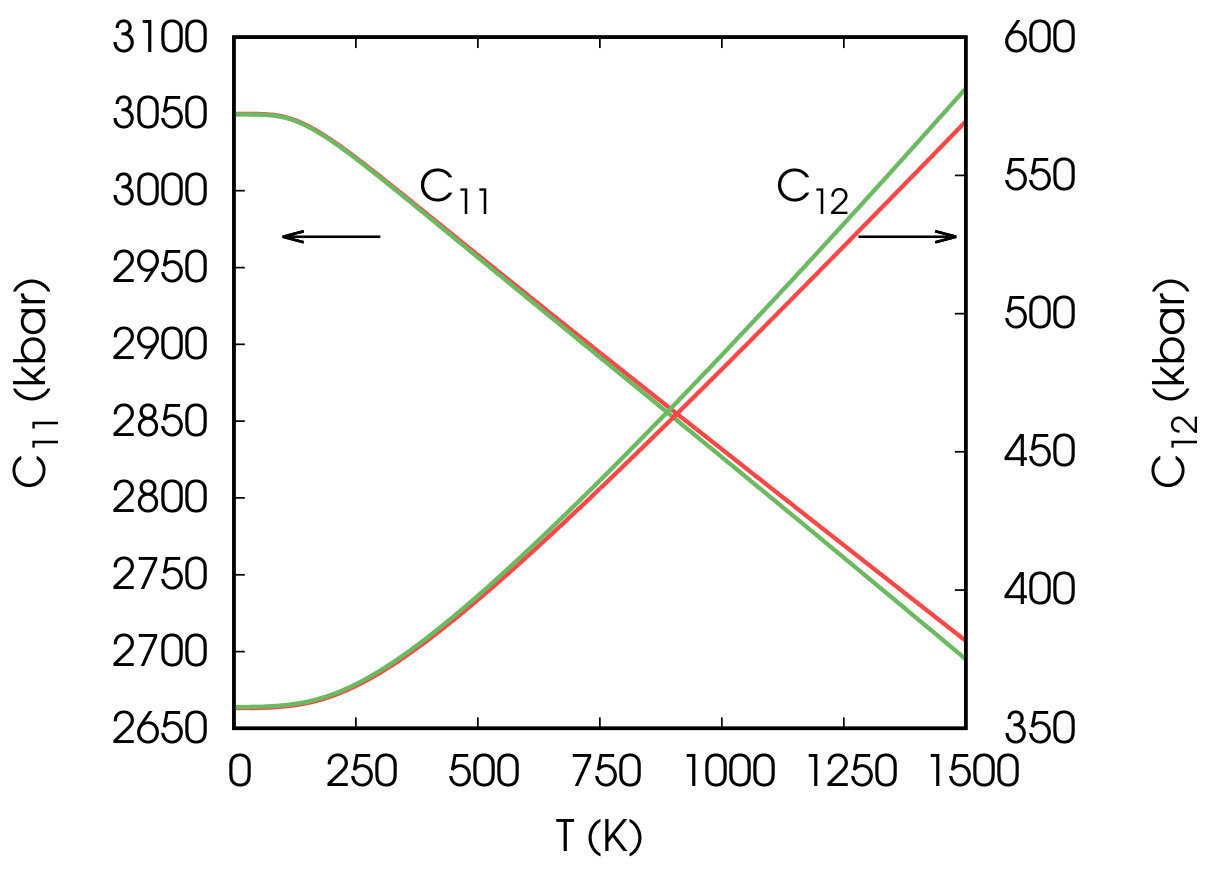}
\caption{Elastic constants $C_{11}$ and $C_{12}$ of Be as a function of temperature calculated as second derivatives of the free energy (within the QHA) at fixed equilibrium geometry. We compare the results obtained with the ZSISA (red lines) and within the FFEM (green lines), a scheme in which the internal $y$ parameter is relaxed at each strain and temperature by minimizing the free energy.}
\label{fig:zsisa}
\end{figure}

\begin{figure}
\centering
\includegraphics[width=\linewidth]{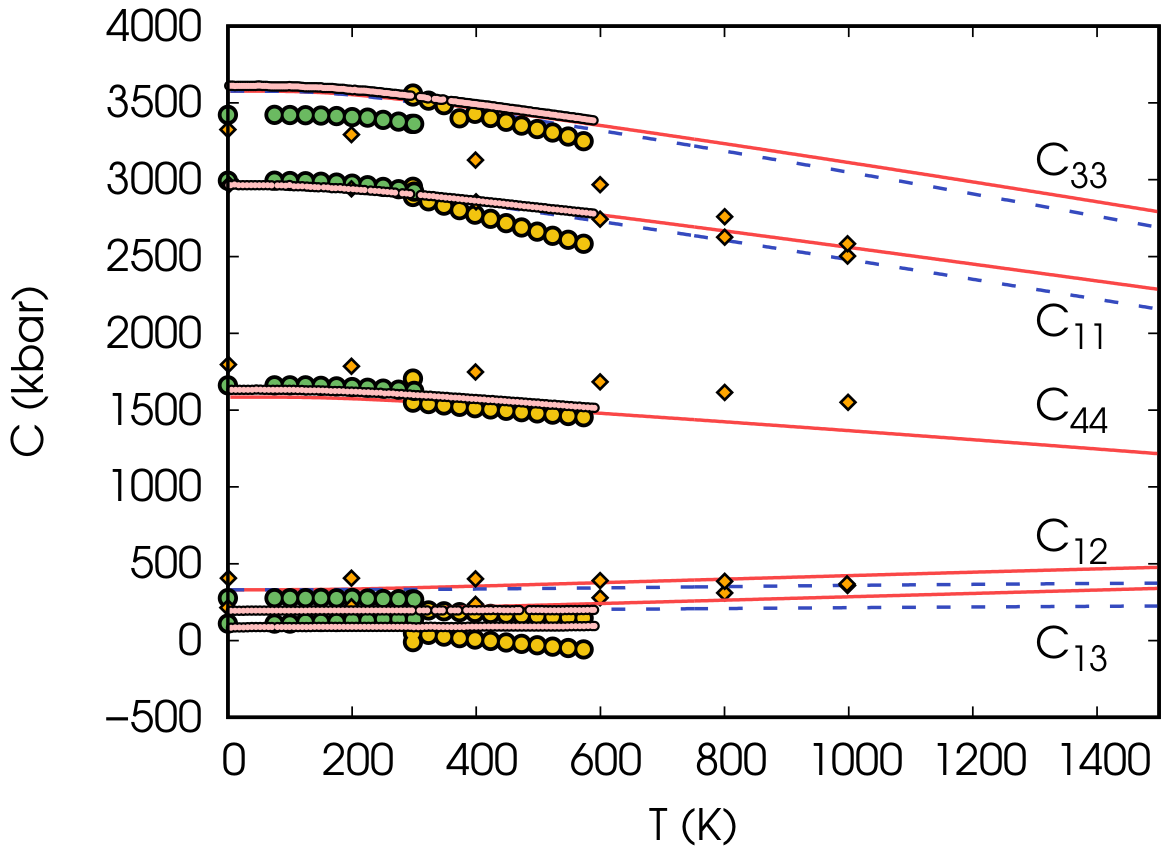}
\caption{Adiabatic elastic constants of Be as a function of temperature (red lines) calculated within the QHA. Atomic relaxations have been dealt with the ZSISA approximation. Calculations have been done only along the ``stress-pressure'' $0$ K isotherm (V-ZSISA). The dots are the experimental points of  Ref.~\cite{rowlands_determination_1972} (yellow dots) and \cite{smith_elastic_1960} (green dots). The diamond are the theoretical QHA results of Ref.~\cite{shao_temperature_2012} while the pink dots are the isothermal QHA elastic constants calculated in Ref.~\cite{robertMultiphaseEquationState2010}. The isothermal elastic constants are also shown (blue dashed lines).}
\label{fig:elastic_qha}
\end{figure}

\begin{figure}
\centering
\includegraphics[width=\linewidth]{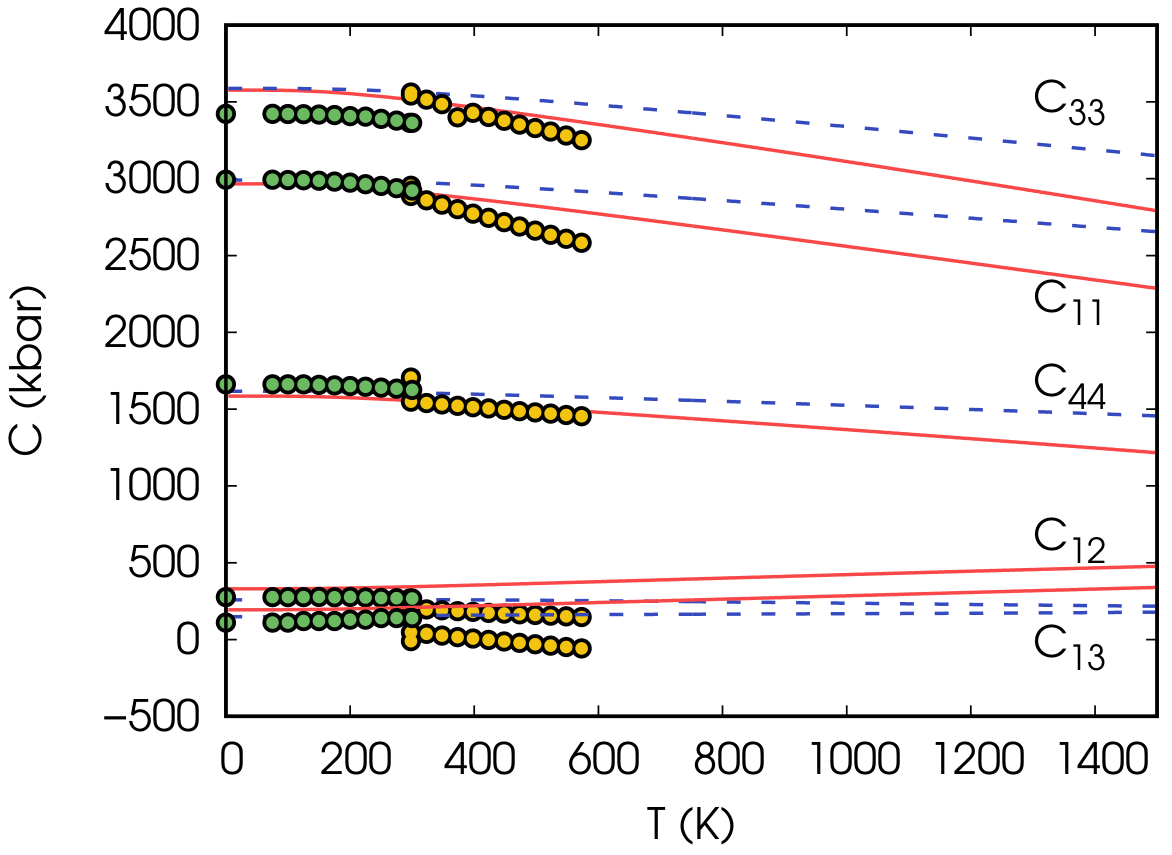}
\caption{Temperature dependent elastic constants of Be as a function of temperature calculated within the V-ZSISA QHA (red lines) are compared with the V-ZSISA QSA
(blue dashed lines). The dots are the experimental points of Ref.~\cite{rowlands_determination_1972} (yellow dots) and \cite{smith_elastic_1960} (green dots). }
\label{fig:elastic_qha_qsa}
\end{figure}

\begin{figure}
\centering
\includegraphics[width=\linewidth]{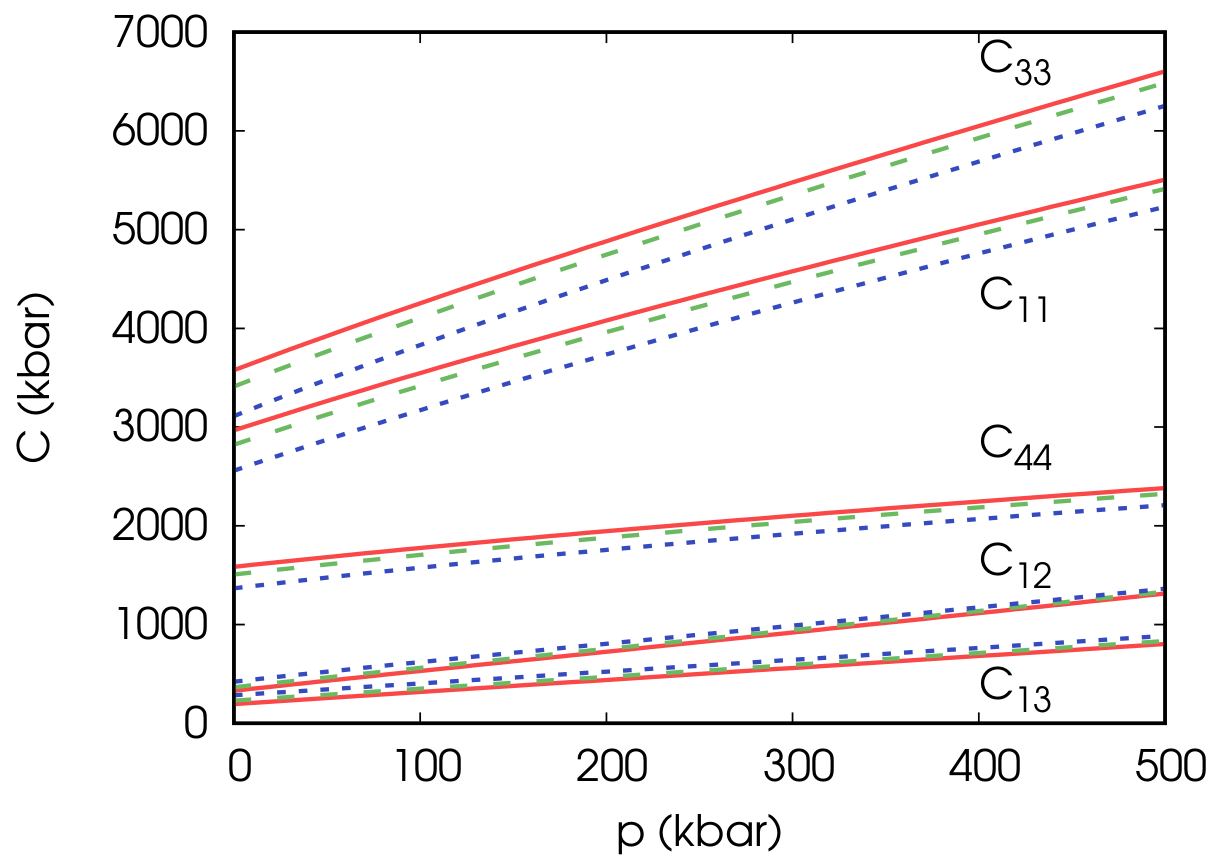}
\caption{Adiabatic pressure dependent elastic constants of Be calculated within the V-ZSISA QHA at three temperatures: $4$ K (red line), $500$ K (green lines) $1000$ K (blue lines). Calculations have been done along the ``stress-pressure'' isotherm.}
\label{fig:elastic_qha_p}
\end{figure}

The equilibrium crystal parameters at $0$ K
obtained from the total energy minimization are reported in Tab.~\ref{tab:elastic} together with our calculated ECs. For comparison,
we also show the ECs of selected
references that are discussed below.
A more complete account of the data available in the literature and of the effects of parameters
such as exchange and correlation energy,
the pseudopotentials, the {\bf k}-point sampling, and the atomic relaxations method is presented in  Ref.~\cite{dal_corso_elastic_2016}.
When compared with the recent experiment
of Ref.~\cite{migliori_berylliums_2004}, our
computed ECs 
at $0$ K match experiment with errors $\Delta C_{11}=138$ kbar ($4 \%$), $\Delta C_{12}=12$ kbar ($4 \%$), $\Delta C_{13}=23$ kbar ($16$ \%), $\Delta C_{33}=107$ kbar ($3$ \%), and $\Delta C_{44}=17$ kbar ($1$ \%). All errors
are within $10 \%$ with the exception of
$C_{13}$, whose value is, however, quite variable also among different experimental reports~\cite{migliori_berylliums_2004}.

The crystal parameters as a function of pressure $a(p)$ and ${c\over a}(p)$ are
calculated from the minimization of the Gibbs energy (Eq.~\ref{gibbs}). The ``stress-pressure'' isotherm at $0$ K is shown in Fig.~\ref{fig:energy} (orange curve) together with the constant energy contours in the plane $a$, $c/a$ and the position of the energy minimum. Pressure
dependent ECs are calculated
in a set of points along this curve. The resulting equation of state
(EOS) and $c/a(p)$ are reported in the supplementary material.

Fig.~\ref{fig:elastic_p} shows the pressure dependent ECs at $0$ K compared with those already published. Our LDA data are in good agreement with the LDA results of Sin'ko et al.~\cite{sinko_relative_2005} available until
$1500$ kbar, and with the PBE ones~\cite{pbe} of Hao et al.~\cite{hao_first-principle_2012} at least until $2000$ kbar. At variance with Ref.~\cite{hao_first-principle_2012} we find no strong deviation from linearity at higher pressures.
The LDA values of $C_{33}$, $C_{13}$, and $C_{44}$ of Luo et al.~\cite{luo_ab_2012}
agree with ours while $C_{11}$ and $C_{12}$ are different. For the latter, better agreement is found by computing the ECs with the ions frozen in their strained positions. 
We mention also that the ECs given in Table II of Ref.~\cite{liu_experimental_2009} are the second derivatives of the total energy with respect to the Lagrangian strains (we call them $\overset{\circ}{C}_{ijkl}$). In
order to compare with our stress-strain results, we have used the following expression~\cite{barron_second-order_1965}:
\begin{equation}
C^T_{ijkl}= \overset{\circ}{C}_{ijkl} + p (\delta_{ij}  \delta_{kl}
-\delta_{ik} \delta_{jl}
-\delta_{il} \delta_{jk}).
\end{equation}
The data of Refs.~\cite{sinko_relative_2005,hao_first-principle_2012}, instead, are the stress-strain ECs and no modification is done. Ref.~\cite{hao_first-principle_2012} uses the PBE functional, so some care should be used to compare with our results.
However, in other materials~\cite{gong_pressure_2024,gong_ab_2024},
we found that on the scale of this figure the differences among functionals are small, and the pressure derivative of the ECs are similar. 

Computing the phonon dispersions on all the points of the two-dimensional grid shown in Fig.~\ref{fig:energy}, we obtain a set of Gibbs energies that can be interpolated with a fourth-degree polynomial whose minimum gives $a(T)$ and ${c\over a}(T)$ at any temperature and pressure.
The ``stress-pressure'' isotherm at $1500$ K is shown in Fig.~\ref{fig:energy}.
In this parameter space, the ``stress-pressure'' isotherms at $0$ K and at $1500$ K are close to each other. This fact is exploited in the literature, where TDECs are calculated only in a few points along the isotherm at $0$ K. This is the so-called V-ZSISA approximation. We estimated the effect of this approximation on the QSA adiabatic ECs. 

Fig.~\ref{fig:elastic_qsa_all_1d} shows two sets of QSA TDECs. In the red curves, the $0$ K ECs are calculated at all points of the two-dimensional grid shown in Fig.~\ref{fig:energy} and interpolated at the crystal parameters that minimize the Gibbs energy. At zero pressure, we interpolate along the green isobar shown in Fig.~\ref{fig:energy} close to the energy minimum. Note that at $0$ K this curve does not start exactly on the energy minimum because of zero-point effects.

In V-ZSISA, instead, the ECs are calculated only in a few points on the ``stress-pressure'' isotherm at $0$ K and interpolated at $a(T)$ for each temperature. Along this line $c/a$ is a function of $a$. At each temperature we use $a(T)$, but $c/a(T)=c/a(a(T))$. 
The change of $c/a$ that one would have moving from the ``stress-pressure'' isotherm at $0$ K to the ``stress-pressure'' isotherm at temperature $T$ is neglected~\cite{kadas_temperature-dependent_2007}.
The results are shown with a blue dashed line in Fig.~\ref{fig:elastic_qsa_all_1d}.
Differences with respect to the complete interpolation are quite small and the temperature dependence is weakly influenced.
The changes from $0$ K to $1500$ K are: 
$\Delta C_{11}=391$ kbar ($13 \%$), $\Delta C_{12}=24$ kbar ($10 \%$), $\Delta C_{13}=-26$ kbar ($-20$ \%), $\Delta C_{33}=457$ kbar ($13$ \%), and $\Delta C_{44}=159$ kbar ($10$ \%)
with the interpolation on the two-dimensional grid and 
$\Delta C_{11}=336$ kbar ($11 \%$), $\Delta C_{12}=41$ kbar ($16\%$), $\Delta C_{13}=-29$ kbar ($-19$ \%), $\Delta C_{33}=438$ kbar ($12$ \%), and $\Delta C_{44}=160$ kbar ($10$ \%) within V-ZSISA. These data agree reasonably well with the QSA calculation of Ref.~\cite{kadas_temperature-dependent_2007} which finds, in the same temperature range,
$\Delta C_{11}=286$ kbar ($10 \%$), $\Delta C_{12}=101$ kbar ($40 \%$), $\Delta C_{33}=368$ kbar ($10$ \%), and $\Delta C_{44}=115$ kbar ($7$ \%). In this reference $C_{13}$ is almost zero and does not change with temperature.

The other approximation that we tested is the ZSISA.  
In Fig.~\ref{fig:zsisa} we show the ECs $C_{11}$ and $C_{12}$ calculated within QHA with and without the ZSISA. These ECs are computed at one reference geometry: the $0$ K crystal parameters (Tab.~\ref{table:2}). 
Hence these ECs have only the contribution of the free energy to the temperature variation. As explained above, 
$C_{11}$ is calculated using the strain $(\epsilon,0,0,0,0,0)$ and is different from the frozen ion value because there is a non-zero internal relaxation, while $C_{12}$ is calculated only later from the strain $(\epsilon,\epsilon,0,0,0,0)$
that does not allow any internal relaxation. It is different from its frozen-ions value because the second derivatives with respect to this strain provide $2C_{11}+2C_{12}$ to which
$C_{11}$ must be subtracted.

For the first (and the third) strain type, we calculate the phonon dispersion in five different atomic positions.
For each equilibrium geometry and strain type, the calculation of these
ECs requires the calculation of the phonon dispersion in $30$ distorted geometries and is therefore much heavier that the ZSISA calculation
that requires only $6$ distorted geometries per strain type. The ZSISA $C_{11}$ is slightly higher than the FFEM, less than $1$ kbar at $4$ K while at $1500$ K the difference are $\Delta C_{11}=-12$ kbar ($-0.4 \%$), $\Delta C_{12}=12$ kbar ($2 \%$, negligible on the scale of the other figures).

Fig.~\ref{fig:elastic_qha} shows the adiabatic QHA TDECs calculated within ZSISA and V-ZSISA. In the same picture, for reference, we show also the isothermal elastic constants. 
From $0$ K to $1500$ K we have the following
decreases $\Delta C_{11}=678$ kbar ($22 \%$), $\Delta C_{12}=-145$ kbar ($-44 \%$), $\Delta C_{13}=-146$ kbar ($-75$ \%), $\Delta C_{33}=784$ ($22$ \%), and $\Delta C_{44}=369$ ($23$ \%). Our data are compared with the QHA results of Ref.~\cite{shao_temperature_2012} (up to $1000$ K) and of Ref.~\cite{robertMultiphaseEquationState2010} (until $600$ K). 
From $0$ K to $600$ K, the temperature dependence predicted by this latter reference agrees very well with our result, although the
values at $0$ K of $C_{12}$ and $C_{13}$ are different from ours. 
Comparing with Ref.~\cite{shao_temperature_2012}
we have a similar temperature dependence for $C_{11}$, $C_{13}$ and $C_{44}$, while
we find a smaller temperature derivative for
$C_{33}$ and a $C_{12}$ that increases with
temperature instead of decreasing.
From $0$ K to $1000$ K  Ref.~\cite{shao_temperature_2012} finds:
$\Delta C_{11}=464$ kbar ($16 \%$), $\Delta C_{12}=37$ kbar ($9 \%$), $\Delta C_{13}=-145$ kbar ($-56$ \%), $\Delta C_{33}=743$ ($22$ \%), and $\Delta C_{44}=206$ ($12$ \%),
to be compared with our adiabatic values:
$\Delta C_{11}=407$ kbar ($13 \%$), $\Delta C_{12}=-92$ kbar ($-27 \%$), $\Delta C_{13}=-91$ kbar ($-46$ \%), $\Delta C_{33}=465$ ($13$ \%), and $\Delta C_{44}=219$ ($13$ \%).

The comparison between the QHA and the QSA elastic constants is shown in Fig.~\ref{fig:elastic_qha_qsa}.  
 The $T=0$ K ECs 
increase with pressure, so we expect a decrease with
temperature that in beryllium expands the volume. Actually this is the picture that one finds in the quasi static approximation (QSA) for the isothermal
ECs.
The QHA $C_{11}$, $C_{33}$, and $C_{44}$ decrease faster with temperature than the QSA ones.
Actually, at fixed structure, QHA $C_{11}$, $C_{33}$, and $C_{44}$ decrease with temperature, and this decrease adds to that due to thermal expansion, the only effect present in the QSA calculation.
Instead the QHA $C_{12}$ and $C_{13}$ both increases with temperature. At fixed geometry, the QHA $C_{12}$ increase and since the thermal expansion causes a decrease as seen for the QSA $C_{12}$, the temperature dependence of the QHA $C_{12}$ is the result of the cancellation of two effects and therefore the sign  might be 
difficult to predict.  Experimental there seem to be a decrease of $C_{12}$ with temperature.
The adiabatic $C_{13}$ increases both within QSA and also within QHA at fixed volume. In the first case this is due to the adiabatic corrections that increase with temperature more than
the decrease of the isothermal $C_{13}$.
So the two increases add up and the QHA
$C_{13}$ increases more than the QSA one.  

Finally, in Fig.~\ref{fig:elastic_qha_p} we report the QHA ECs as a function of pressure at $4$ K, $500$ K, and $1000$ K. In the pressure range from $0$ kbar to $500$ kbar, shown in the figure, the nonlinearities are small. There is no previous information on these elastic constants and we hope that the present calculation will stimulate their measurement at high temperature and pressure, together with a reassessment of the zero pressure high temperature behaviour.

\section{Conclusions}
We presented the QHA TDECs of beryllium calculated (within the V-ZSISA) in eight reference geometries along the ``stress-pressure'' $0$ K isotherm and interpolated at $a(T)$. For $C_{11}$ and $C_{12}$, atomic relaxations have been dealt mainly within the ZSISA approximation. We have verified using the QSA that the `stress-pressure'' $0$ K isotherm interpolation (V-ZSISA) gives results close to the interpolation made along the $0$ kbar isobar. Moreover, we have compared the ZSISA approximation with the full free energy minimization (FFEM) with respect to the atomic positions, finding that for the present case ZSISA is a very good approximation.
Comparison of our results with previous QSA and QHA calculations shows substantial agreement especially with Ref.~\cite{kadas_temperature-dependent_2007} for
the QSA and Ref.~\cite{robertMultiphaseEquationState2010} for the QHA.
Moreover, we provided the first estimate of the pressure dependent (up to $500$ kbar) elastic constants at temperature of $500$ K and $1000$ K. We hope that these calculations will stimulate and
support an experimental investigation of these quantities that are still unknown in beryllium.

The plots of the thermal expansion and of the isobaric heat capacity in the supplemental 
material~\cite{supplemental} (see also references 
\cite{gordon_high_1949,nakanoXrayDiffractionStudy2002,song_modified_2007,lazzeriAbinitioDynamicalProperties1998,robertEquationStateElastic2006} therein)
show that QHA might be a reasonable approximation until $800$ K where the QHA is able to reproduce the experimental results. In general QHA is expected to be accurate until $2/3$ of the melting temperature so our data might require corrections above $1000$ K~\cite{allenAnharmonicPhononQuasiparticle2015}, even if we have plotted them until $1500$ K.

The calculations performed here of the QHA TCECs, required the phonon dispersion on $8 \times 30=240$ geometries ( $30$ distorted configurations of $8$ equilibrium geometries).
With much more effort, slightly more accurate calculations could have been done by computing the quasi-harmonic elastic constants taking as reference geometries all the two-dimensional mesh of $a$ and $c/a$ parameters. This calculation would require the phonon dispersions in $14\times 7 \times 30 =2940$ geometries ($30$ distorted configuration of a grid $14\times 7$ of equilibrium geometries) and is presently beyond our computational resources, but it could become feasible soon. Presently, beryllium does not seem to require such an effort, but we have presented a workflow capable of going beyond both the V-ZSISA and ZSISA when necessary and it might be interesting to see if the conclusions reached in beryllium remain valid also for the other hcp metals. 
All methods used in this paper have been implemented in the \texttt{thermo\_pw} software~\cite{dal_corso_thermo_pw_2022} and
are publicly available. 

\begin{acknowledgments}
    
This work has been supported by the Italian MUR (Ministry of University and Research) through the National Centre for HPC, Big Data, and Quantum Computing (grant No. CN00000013). The SISSA has provided computational facilities through its Linux Cluster, ITCS, and the SISSA-CINECA 2021-2024 Agreement.  Partial support has been received from the European Union through the MAX ``MAterials design at the eXascale" Centre of Excellence for Supercomputing applications (Grant agreement No. 101093374, co-funded by the European High Performance Computing joint Undertaking (JU) and participating countries 824143).  

\end{acknowledgments}


\bibliographystyle{apsrev4-1}
\bibliography{apssamp}

\end{document}

%% file: table2.tex
\caption{\label{table:2} The $0$ K elastic constants  compared with experiment and previous calculations. $B$, $E$, $G$, and $\nu$ are the bulk modulus, the Young's modulus, the shear modulus, and the Poisson's ratio, of polycrystalline beryllium calculated within the Voigt-Reuss-Hill approximation, respectively.}
\begin{ruledtabular}
\begin{tabular}{ccccccccccccc}
 & T &$a_0$ & $\frac{a_0}{c_0}$  & $C_{11}$  & $C_{12}$ & $C_{13}$ & $C_{33}$ & $C_{44}$ & $B$ & $E$ & $G$ & ${\nu}$ \\
  & (K) & (a.u.) &  & (kbar) & (kbar) & (kbar) & (kbar) & (kbar) & (kbar) & (kbar) & (kbar) & \\
\hline
This study (LDA) & 0 & 4.244 & 1.573 & 3074 & 280  & 163 & 3674 & 1639 &  1223 & 3259 & 1543 & 0.06\\
   Ref.~\cite{sinko_relative_2005} & 0 & 4.281 & 1.573 &  3008 & 141  & 71 & 3595 & 1602 & 1127  & 3182& 1545 & 0.06 \\
     Ref.~\cite{hao_first-principle_2012}(LDA) & 0 & 4.312 & 1.567 &  2150 & 610  & -60 & 3500 & 1560 & 970 & 2532 & 1114 & 0.06 \\
    Ref.~\cite{luo_ab_2012}(LDA) & 0 & 4.248 & 1.57 &  3109 & 195  & 191 & 3595 & 1621 & 1215 & 3267 &1552 & 0.05 \\  Ref.~\cite{kadas_temperature-dependent_2007}(PBE) & 0 &  & 1.577 & 2882  & 254  & 0 & 3652 & 1567 & 1100 &  3080 & 1490 & 0.03 \\   
Ref.~\cite{robertMultiphaseEquationState2010}(PBE) & 0 &  & 1.575 & 2965  & 194  & 83 & 3612 & 1632 & 1137 & 3177  & 1536 & 0.03 \\
Ref.~\cite{shao_temperature_2012}(LDA) &  &  &  &  2966 & 403  & 209 & 3323 & 1798 & 1210 & 3214& 1520 & 0.06 \\
     Ref.~\cite{migliori_berylliums_2004} (Expt.) &  & 4.319\footnotemark[1] & 1.568\footnotemark[1] &  2936 & 268  & 140 & 3567 & 1622 & 1168 & 3152 & 1501 & 0.05\\
\end{tabular}
\end{ruledtabular}
\label{tab:elastic}
\footnotetext[1]{Ref.~\cite{mackayLatticeParameterHardness1963}.}

%% file: paper_clean.bbl
\begin{thebibliography}{52}%
\makeatletter
\providecommand \@ifxundefined [1]{%
 \@ifx{#1\undefined}
}%
\providecommand \@ifnum [1]{%
 \ifnum #1\expandafter \@firstoftwo
 \else \expandafter \@secondoftwo
 \fi
}%
\providecommand \@ifx [1]{%
 \ifx #1\expandafter \@firstoftwo
 \else \expandafter \@secondoftwo
 \fi
}%
\providecommand \natexlab [1]{#1}%
\providecommand \enquote  [1]{``#1''}%
\providecommand \bibnamefont  [1]{#1}%
\providecommand \bibfnamefont [1]{#1}%
\providecommand \citenamefont [1]{#1}%
\providecommand \href@noop [0]{\@secondoftwo}%
\providecommand \href [0]{\begingroup \@sanitize@url \@href}%
\providecommand \@href[1]{\@@startlink{#1}\@@href}%
\providecommand \@@href[1]{\endgroup#1\@@endlink}%
\providecommand \@sanitize@url [0]{\catcode `\\12\catcode `\$12\catcode `\&12\catcode `\#12\catcode `\^12\catcode `\_12\catcode `\%12\relax}%
\providecommand \@@startlink[1]{}%
\providecommand \@@endlink[0]{}%
\providecommand \url  [0]{\begingroup\@sanitize@url \@url }%
\providecommand \@url [1]{\endgroup\@href {#1}{\urlprefix }}%
\providecommand \urlprefix  [0]{URL }%
\providecommand \Eprint [0]{\href }%
\providecommand \doibase [0]{https://doi.org/}%
\providecommand \selectlanguage [0]{\@gobble}%
\providecommand \bibinfo  [0]{\@secondoftwo}%
\providecommand \bibfield  [0]{\@secondoftwo}%
\providecommand \translation [1]{[#1]}%
\providecommand \BibitemOpen [0]{}%
\providecommand \bibitemStop [0]{}%
\providecommand \bibitemNoStop [0]{.\EOS\space}%
\providecommand \EOS [0]{\spacefactor3000\relax}%
\providecommand \BibitemShut  [1]{\csname bibitem#1\endcsname}%
\let\auto@bib@innerbib\@empty
\bibitem [{\citenamefont {Arblaster}(2016)}]{arblaster_thermodynamic_2016}%
  \BibitemOpen
  \bibfield  {author} {\bibinfo {author} {\bibfnamefont {J.~W.}\ \bibnamefont {Arblaster}},\ }\bibfield  {title} {\bibinfo {title} {Thermodynamic properties of beryllium},\ }\href {https://doi.org/10.1007/s11669-016-0488-5} {\bibfield  {journal} {\bibinfo  {journal} {Journal of Phase Equilibria and Diffusion}\ }\textbf {\bibinfo {volume} {37}},\ \bibinfo {pages} {581} (\bibinfo {year} {2016})}\BibitemShut {NoStop}%
\bibitem [{\citenamefont {Bodryakov}(2014)}]{bodryakov_correlation_2014}%
  \BibitemOpen
  \bibfield  {author} {\bibinfo {author} {\bibfnamefont {V.~Y.}\ \bibnamefont {Bodryakov}},\ }\bibfield  {title} {\bibinfo {title} {Correlation of temperature dependencies of thermal expansion and heat capacity of refractory metal up to the melting point: Molybdenum},\ }\href {https://doi.org/10.1134/S0018151X14040051} {\bibfield  {journal} {\bibinfo  {journal} {High Temperature}\ }\textbf {\bibinfo {volume} {52}},\ \bibinfo {pages} {840} (\bibinfo {year} {2014})}\BibitemShut {NoStop}%
\bibitem [{\citenamefont {Stedman}\ \emph {et~al.}(1976)\citenamefont {Stedman}, \citenamefont {Amilius}, \citenamefont {Pauli},\ and\ \citenamefont {Sundin}}]{stedman_phonon_1976}%
  \BibitemOpen
  \bibfield  {author} {\bibinfo {author} {\bibfnamefont {R.}~\bibnamefont {Stedman}}, \bibinfo {author} {\bibfnamefont {Z.}~\bibnamefont {Amilius}}, \bibinfo {author} {\bibfnamefont {R.}~\bibnamefont {Pauli}},\ and\ \bibinfo {author} {\bibfnamefont {O.}~\bibnamefont {Sundin}},\ }\bibfield  {title} {\bibinfo {title} {Phonon spectrum of beryllium at 80k},\ }\href {https://doi.org/10.1088/0305-4608/6/2/012} {\bibfield  {journal} {\bibinfo  {journal} {Journal of Physics F: Metal Physics}\ }\textbf {\bibinfo {volume} {6}},\ \bibinfo {pages} {157} (\bibinfo {year} {1976})}\BibitemShut {NoStop}%
\bibitem [{\citenamefont {Lazicki}\ \emph {et~al.}(2012)\citenamefont {Lazicki}, \citenamefont {Dewaele}, \citenamefont {Loubeyre},\ and\ \citenamefont {Mezouar}}]{lazicki_high-pressure--temperature_2012}%
  \BibitemOpen
  \bibfield  {author} {\bibinfo {author} {\bibfnamefont {A.}~\bibnamefont {Lazicki}}, \bibinfo {author} {\bibfnamefont {A.}~\bibnamefont {Dewaele}}, \bibinfo {author} {\bibfnamefont {P.}~\bibnamefont {Loubeyre}},\ and\ \bibinfo {author} {\bibfnamefont {M.}~\bibnamefont {Mezouar}},\ }\bibfield  {title} {\bibinfo {title} {High-pressure--temperature phase diagram and the equation of state of beryllium},\ }\href {https://doi.org/10.1103/PhysRevB.86.174118} {\bibfield  {journal} {\bibinfo  {journal} {Physical Review B}\ }\textbf {\bibinfo {volume} {86}},\ \bibinfo {pages} {174118} (\bibinfo {year} {2012})}\BibitemShut {NoStop}%
\bibitem [{\citenamefont {Luo}\ \emph {et~al.}(2012)\citenamefont {Luo}, \citenamefont {Cai}, \citenamefont {Chen}, \citenamefont {Jing},\ and\ \citenamefont {Alfè}}]{luo_ab_2012}%
  \BibitemOpen
  \bibfield  {author} {\bibinfo {author} {\bibfnamefont {F.}~\bibnamefont {Luo}}, \bibinfo {author} {\bibfnamefont {L.-C.}\ \bibnamefont {Cai}}, \bibinfo {author} {\bibfnamefont {X.-R.}\ \bibnamefont {Chen}}, \bibinfo {author} {\bibfnamefont {F.-Q.}\ \bibnamefont {Jing}},\ and\ \bibinfo {author} {\bibfnamefont {D.}~\bibnamefont {Alfè}},\ }\bibfield  {title} {\bibinfo {title} {Ab initio calculation of lattice dynamics and thermodynamic properties of beryllium},\ }\href {https://doi.org/10.1063/1.3688344} {\bibfield  {journal} {\bibinfo  {journal} {Journal of Applied Physics}\ }\textbf {\bibinfo {volume} {111}},\ \bibinfo {pages} {053503} (\bibinfo {year} {2012})}\BibitemShut {NoStop}%
\bibitem [{\citenamefont {Hao}\ and\ \citenamefont {Zhu}(2012)}]{hao_first-principle_2012}%
  \BibitemOpen
  \bibfield  {author} {\bibinfo {author} {\bibfnamefont {A.}~\bibnamefont {Hao}}\ and\ \bibinfo {author} {\bibfnamefont {Y.}~\bibnamefont {Zhu}},\ }\bibfield  {title} {\bibinfo {title} {First-principle investigations of structural stability of beryllium under high pressure},\ }\href {https://doi.org/10.1063/1.4739615} {\bibfield  {journal} {\bibinfo  {journal} {Journal of Applied Physics}\ }\textbf {\bibinfo {volume} {112}},\ \bibinfo {pages} {023519} (\bibinfo {year} {2012})}\BibitemShut {NoStop}%
\bibitem [{\citenamefont {Kádas}\ \emph {et~al.}(2007)\citenamefont {Kádas}, \citenamefont {Vitos}, \citenamefont {Ahuja}, \citenamefont {Johansson},\ and\ \citenamefont {Kollár}}]{kadas_temperature-dependent_2007}%
  \BibitemOpen
  \bibfield  {author} {\bibinfo {author} {\bibfnamefont {K.}~\bibnamefont {Kádas}}, \bibinfo {author} {\bibfnamefont {L.}~\bibnamefont {Vitos}}, \bibinfo {author} {\bibfnamefont {R.}~\bibnamefont {Ahuja}}, \bibinfo {author} {\bibfnamefont {B.}~\bibnamefont {Johansson}},\ and\ \bibinfo {author} {\bibfnamefont {J.}~\bibnamefont {Kollár}},\ }\bibfield  {title} {\bibinfo {title} {Temperature-dependent elastic properties of \${\textbackslash}ensuremath\{{\textbackslash}alpha\}\$-beryllium from first principles},\ }\href {https://doi.org/10.1103/PhysRevB.76.235109} {\bibfield  {journal} {\bibinfo  {journal} {Physical Review B}\ }\textbf {\bibinfo {volume} {76}},\ \bibinfo {pages} {235109} (\bibinfo {year} {2007})}\BibitemShut {NoStop}%
\bibitem [{\citenamefont {Sin’ko}\ and\ \citenamefont {Smirnov}(2005)}]{sinko_relative_2005}%
  \BibitemOpen
  \bibfield  {author} {\bibinfo {author} {\bibfnamefont {G.~V.}\ \bibnamefont {Sin’ko}}\ and\ \bibinfo {author} {\bibfnamefont {N.~A.}\ \bibnamefont {Smirnov}},\ }\bibfield  {title} {\bibinfo {title} {Relative stability and elastic properties of hcp, bcc, and fcc beryllium under pressure},\ }\href {https://doi.org/10.1103/PhysRevB.71.214108} {\bibfield  {journal} {\bibinfo  {journal} {Physical Review B}\ }\textbf {\bibinfo {volume} {71}},\ \bibinfo {pages} {214108} (\bibinfo {year} {2005})}\BibitemShut {NoStop}%
\bibitem [{\citenamefont {Robert}\ \emph {et~al.}(2010)\citenamefont {Robert}, \citenamefont {Legrand},\ and\ \citenamefont {Bernard}}]{robertMultiphaseEquationState2010}%
  \BibitemOpen
  \bibfield  {author} {\bibinfo {author} {\bibfnamefont {G.}~\bibnamefont {Robert}}, \bibinfo {author} {\bibfnamefont {P.}~\bibnamefont {Legrand}},\ and\ \bibinfo {author} {\bibfnamefont {S.}~\bibnamefont {Bernard}},\ }\bibfield  {title} {\bibinfo {title} {Multiphase equation of state and elastic moduli of solid beryllium from first principles},\ }\href {https://doi.org/10.1103/PhysRevB.82.104118} {\bibfield  {journal} {\bibinfo  {journal} {Physical Review B}\ }\textbf {\bibinfo {volume} {82}},\ \bibinfo {pages} {104118} (\bibinfo {year} {2010})}\BibitemShut {NoStop}%
\bibitem [{\citenamefont {Shao}\ \emph {et~al.}(2012)\citenamefont {Shao}, \citenamefont {Wen}, \citenamefont {Melnik}, \citenamefont {Yao}, \citenamefont {Kawazoe},\ and\ \citenamefont {Tian}}]{shao_temperature_2012}%
  \BibitemOpen
  \bibfield  {author} {\bibinfo {author} {\bibfnamefont {T.}~\bibnamefont {Shao}}, \bibinfo {author} {\bibfnamefont {B.}~\bibnamefont {Wen}}, \bibinfo {author} {\bibfnamefont {R.}~\bibnamefont {Melnik}}, \bibinfo {author} {\bibfnamefont {S.}~\bibnamefont {Yao}}, \bibinfo {author} {\bibfnamefont {Y.}~\bibnamefont {Kawazoe}},\ and\ \bibinfo {author} {\bibfnamefont {Y.}~\bibnamefont {Tian}},\ }\bibfield  {title} {\bibinfo {title} {Temperature dependent elastic constants for crystals with arbitrary symmetry: Combined first principles and continuum elasticity theory},\ }\href {https://doi.org/10.1063/1.4704698} {\bibfield  {journal} {\bibinfo  {journal} {Journal of Applied Physics}\ }\textbf {\bibinfo {volume} {111}},\ \bibinfo {pages} {083525} (\bibinfo {year} {2012})}\BibitemShut {NoStop}%
\bibitem [{\citenamefont {Wu}\ \emph {et~al.}(2021)\citenamefont {Wu}, \citenamefont {González-Cataldo},\ and\ \citenamefont {Militzer}}]{wu_high-pressure_2021}%
  \BibitemOpen
  \bibfield  {author} {\bibinfo {author} {\bibfnamefont {J.}~\bibnamefont {Wu}}, \bibinfo {author} {\bibfnamefont {F.}~\bibnamefont {González-Cataldo}},\ and\ \bibinfo {author} {\bibfnamefont {B.}~\bibnamefont {Militzer}},\ }\bibfield  {title} {\bibinfo {title} {High-pressure phase diagram of beryllium from ab initio free-energy calculations},\ }\href {https://doi.org/10.1103/PhysRevB.104.014103} {\bibfield  {journal} {\bibinfo  {journal} {Physical Review B}\ }\textbf {\bibinfo {volume} {104}},\ \bibinfo {pages} {014103} (\bibinfo {year} {2021})}\BibitemShut {NoStop}%
\bibitem [{\citenamefont {Migliori}\ \emph {et~al.}(2004)\citenamefont {Migliori}, \citenamefont {Ledbetter}, \citenamefont {Thoma},\ and\ \citenamefont {Darling}}]{migliori_berylliums_2004}%
  \BibitemOpen
  \bibfield  {author} {\bibinfo {author} {\bibfnamefont {A.}~\bibnamefont {Migliori}}, \bibinfo {author} {\bibfnamefont {H.}~\bibnamefont {Ledbetter}}, \bibinfo {author} {\bibfnamefont {D.~J.}\ \bibnamefont {Thoma}},\ and\ \bibinfo {author} {\bibfnamefont {T.~W.}\ \bibnamefont {Darling}},\ }\bibfield  {title} {\bibinfo {title} {Beryllium’s monocrystal and polycrystal elastic constants},\ }\href {https://doi.org/10.1063/1.1644633} {\bibfield  {journal} {\bibinfo  {journal} {Journal of Applied Physics}\ }\textbf {\bibinfo {volume} {95}},\ \bibinfo {pages} {2436} (\bibinfo {year} {2004})}\BibitemShut {NoStop}%
\bibitem [{\citenamefont {Dal~Corso}(2016)}]{dal_corso_elastic_2016}%
  \BibitemOpen
  \bibfield  {author} {\bibinfo {author} {\bibfnamefont {A.}~\bibnamefont {Dal~Corso}},\ }\bibfield  {title} {\bibinfo {title} {Elastic constants of beryllium: a first-principles investigation},\ }\href@noop {} {\bibfield  {journal} {\bibinfo  {journal} {Journal of Physics: Condensed Matter}\ }\textbf {\bibinfo {volume} {28}},\ \bibinfo {pages} {075401} (\bibinfo {year} {2016})}\BibitemShut {NoStop}%
\bibitem [{\citenamefont {Smith}\ and\ \citenamefont {Arbogast}(1960)}]{smith_elastic_1960}%
  \BibitemOpen
  \bibfield  {author} {\bibinfo {author} {\bibfnamefont {J.~F.}\ \bibnamefont {Smith}}\ and\ \bibinfo {author} {\bibfnamefont {C.~L.}\ \bibnamefont {Arbogast}},\ }\bibfield  {title} {\bibinfo {title} {Elastic constants of single crystal beryllium},\ }\href {https://doi.org/10.1063/1.1735427} {\bibfield  {journal} {\bibinfo  {journal} {Journal of Applied Physics}\ }\textbf {\bibinfo {volume} {31}},\ \bibinfo {pages} {99} (\bibinfo {year} {1960})}\BibitemShut {NoStop}%
\bibitem [{\citenamefont {Rowlands}\ and\ \citenamefont {White}(1972)}]{rowlands_determination_1972}%
  \BibitemOpen
  \bibfield  {author} {\bibinfo {author} {\bibfnamefont {W.~D.}\ \bibnamefont {Rowlands}}\ and\ \bibinfo {author} {\bibfnamefont {J.~S.}\ \bibnamefont {White}},\ }\bibfield  {title} {\bibinfo {title} {The determination of the elastic constants of beryllium in the temperature range 25 to 300 c},\ }\href {https://doi.org/10.1088/0305-4608/2/2/011} {\bibfield  {journal} {\bibinfo  {journal} {Journal of Physics F: Metal Physics}\ }\textbf {\bibinfo {volume} {2}},\ \bibinfo {pages} {231} (\bibinfo {year} {1972})}\BibitemShut {NoStop}%
\bibitem [{\citenamefont {Nadal}\ and\ \citenamefont {Bourgeois}(2010)}]{nadalElasticModuliBeryllium2010}%
  \BibitemOpen
  \bibfield  {author} {\bibinfo {author} {\bibfnamefont {M.-H.}\ \bibnamefont {Nadal}}\ and\ \bibinfo {author} {\bibfnamefont {L.}~\bibnamefont {Bourgeois}},\ }\bibfield  {title} {\bibinfo {title} {Elastic moduli of beryllium versus temperature: {{Experimental}} data updating},\ }\href {https://doi.org/10.1063/1.3455859} {\bibfield  {journal} {\bibinfo  {journal} {Journal of Applied Physics}\ }\textbf {\bibinfo {volume} {108}},\ \bibinfo {pages} {033512} (\bibinfo {year} {2010})}\BibitemShut {NoStop}%
\bibitem [{\citenamefont {Allan}\ \emph {et~al.}(1996)\citenamefont {Allan}, \citenamefont {Barron},\ and\ \citenamefont {Bruno}}]{allanZeroStaticInternal1996a}%
  \BibitemOpen
  \bibfield  {author} {\bibinfo {author} {\bibfnamefont {N.~L.}\ \bibnamefont {Allan}}, \bibinfo {author} {\bibfnamefont {T.~H.~K.}\ \bibnamefont {Barron}},\ and\ \bibinfo {author} {\bibfnamefont {J.~A.~O.}\ \bibnamefont {Bruno}},\ }\bibfield  {title} {\bibinfo {title} {The zero static internal stress approximation in lattice dynamics, and the calculation of isotope effects on molar volumes},\ }\href {https://doi.org/10.1063/1.472684} {\bibfield  {journal} {\bibinfo  {journal} {The Journal of Chemical Physics}\ }\textbf {\bibinfo {volume} {105}},\ \bibinfo {pages} {8300} (\bibinfo {year} {1996})}\BibitemShut {NoStop}%
\bibitem [{\citenamefont {Masuki}\ \emph {et~al.}(2023)\citenamefont {Masuki}, \citenamefont {Nomoto}, \citenamefont {Arita},\ and\ \citenamefont {Tadano}}]{masukiFullOptimizationQuasiharmonic2023a}%
  \BibitemOpen
  \bibfield  {author} {\bibinfo {author} {\bibfnamefont {R.}~\bibnamefont {Masuki}}, \bibinfo {author} {\bibfnamefont {T.}~\bibnamefont {Nomoto}}, \bibinfo {author} {\bibfnamefont {R.}~\bibnamefont {Arita}},\ and\ \bibinfo {author} {\bibfnamefont {T.}~\bibnamefont {Tadano}},\ }\bibfield  {title} {\bibinfo {title} {Full optimization of quasiharmonic free energy with an anharmonic lattice model: {{Application}} to thermal expansion and pyroelectricity of wurtzite {{GaN}} and {{ZnO}}},\ }\href {https://doi.org/10.1103/PhysRevB.107.134119} {\bibfield  {journal} {\bibinfo  {journal} {Physical Review B}\ }\textbf {\bibinfo {volume} {107}},\ \bibinfo {pages} {134119} (\bibinfo {year} {2023})}\BibitemShut {NoStop}%
\bibitem [{\citenamefont {Carrier}\ \emph {et~al.}(2007)\citenamefont {Carrier}, \citenamefont {Wentzcovitch},\ and\ \citenamefont {Tsuchiya}}]{carrierFirstprinciplesPredictionCrystal2007}%
  \BibitemOpen
  \bibfield  {author} {\bibinfo {author} {\bibfnamefont {P.}~\bibnamefont {Carrier}}, \bibinfo {author} {\bibfnamefont {R.}~\bibnamefont {Wentzcovitch}},\ and\ \bibinfo {author} {\bibfnamefont {J.}~\bibnamefont {Tsuchiya}},\ }\bibfield  {title} {\bibinfo {title} {First-principles prediction of crystal structures at high temperatures using the quasiharmonic approximation},\ }\href {https://doi.org/10.1103/PhysRevB.76.064116} {\bibfield  {journal} {\bibinfo  {journal} {Physical Review B}\ }\textbf {\bibinfo {volume} {76}},\ \bibinfo {pages} {064116} (\bibinfo {year} {2007})}\BibitemShut {NoStop}%
\bibitem [{\citenamefont {Malica}\ and\ \citenamefont {Dal~Corso}(2020{\natexlab{a}})}]{malica_quasi-harmonic_2020}%
  \BibitemOpen
  \bibfield  {author} {\bibinfo {author} {\bibfnamefont {C.}~\bibnamefont {Malica}}\ and\ \bibinfo {author} {\bibfnamefont {A.}~\bibnamefont {Dal~Corso}},\ }\bibfield  {title} {\bibinfo {title} {Quasi-harmonic temperature dependent elastic constants: applications to silicon, aluminum, and silver},\ }\href@noop {} {\bibfield  {journal} {\bibinfo  {journal} {Journal of Physics: Condensed Matter}\ }\textbf {\bibinfo {volume} {32}},\ \bibinfo {pages} {315902} (\bibinfo {year} {2020}{\natexlab{a}})}\BibitemShut {NoStop}%
\bibitem [{\citenamefont {Malica}\ and\ \citenamefont {Dal~Corso}(2021)}]{malica_quasi-harmonic_2021}%
  \BibitemOpen
  \bibfield  {author} {\bibinfo {author} {\bibfnamefont {C.}~\bibnamefont {Malica}}\ and\ \bibinfo {author} {\bibfnamefont {A.}~\bibnamefont {Dal~Corso}},\ }\bibfield  {title} {\bibinfo {title} {Quasi-harmonic thermoelasticity of palladium, platinum, copper, and gold from first principles},\ }\href@noop {} {\bibfield  {journal} {\bibinfo  {journal} {Journal of Physics: Condensed Matter}\ }\textbf {\bibinfo {volume} {33}},\ \bibinfo {pages} {475901} (\bibinfo {year} {2021})}\BibitemShut {NoStop}%
\bibitem [{\citenamefont {Gong}\ and\ \citenamefont {Dal~Corso}(2024{\natexlab{a}})}]{gong_pressure_2024}%
  \BibitemOpen
  \bibfield  {author} {\bibinfo {author} {\bibfnamefont {X.}~\bibnamefont {Gong}}\ and\ \bibinfo {author} {\bibfnamefont {A.}~\bibnamefont {Dal~Corso}},\ }\bibfield  {title} {\bibinfo {title} {Pressure and temperature dependent ab-initio quasi-harmonic thermoelastic properties of tungsten},\ }\href {https://doi.org/10.1088/1361-648X/ad3ac3} {\bibfield  {journal} {\bibinfo  {journal} {Journal of Physics: Condensed Matter}\ }\textbf {\bibinfo {volume} {36}},\ \bibinfo {pages} {285702} (\bibinfo {year} {2024}{\natexlab{a}})}\BibitemShut {NoStop}%
\bibitem [{\citenamefont {Gong}\ and\ \citenamefont {Dal~Corso}(2024{\natexlab{b}})}]{gong_ab_2024}%
  \BibitemOpen
  \bibfield  {author} {\bibinfo {author} {\bibfnamefont {X.}~\bibnamefont {Gong}}\ and\ \bibinfo {author} {\bibfnamefont {A.}~\bibnamefont {Dal~Corso}},\ }\bibfield  {title} {\bibinfo {title} {Ab initio quasi-harmonic thermoelasticity of molybdenum at high temperature and pressure},\ }\href {https://doi.org/10.1063/5.0212162} {\bibfield  {journal} {\bibinfo  {journal} {The Journal of Chemical Physics}\ }\textbf {\bibinfo {volume} {160}},\ \bibinfo {pages} {244703} (\bibinfo {year} {2024}{\natexlab{b}})}\BibitemShut {NoStop}%
\bibitem [{dal(2014)}]{dal_corso_thermo_pw_2022}%
  \BibitemOpen
  \href@noop {} {\bibinfo {title} {\texttt{{thermo}\_pw}}},\ \bibinfo {howpublished} {can be found at the webpage https://github.com/dalcorso/thermo\_pw} (\bibinfo {year} {2014})\BibitemShut {NoStop}%
\bibitem [{\citenamefont {Palumbo}\ and\ \citenamefont {Dal~Corso}(2017{\natexlab{a}})}]{palumboLatticeDynamicsThermophysical2017c}%
  \BibitemOpen
  \bibfield  {author} {\bibinfo {author} {\bibfnamefont {M.}~\bibnamefont {Palumbo}}\ and\ \bibinfo {author} {\bibfnamefont {A.}~\bibnamefont {Dal~Corso}},\ }\bibfield  {title} {\bibinfo {title} {Lattice dynamics and thermophysical properties of h.c.p. {{Os}} and {{Ru}} from the quasi-harmonic approximation},\ }\href {https://doi.org/10.1088/1361-648X/aa7dca} {\bibfield  {journal} {\bibinfo  {journal} {Journal of Physics: Condensed Matter}\ }\textbf {\bibinfo {volume} {29}},\ \bibinfo {pages} {395401} (\bibinfo {year} {2017}{\natexlab{a}})}\BibitemShut {NoStop}%
\bibitem [{\citenamefont {Palumbo}\ and\ \citenamefont {Dal~Corso}(2017{\natexlab{b}})}]{pal2}%
  \BibitemOpen
  \bibfield  {author} {\bibinfo {author} {\bibfnamefont {M.}~\bibnamefont {Palumbo}}\ and\ \bibinfo {author} {\bibfnamefont {A.}~\bibnamefont {Dal~Corso}},\ }\bibfield  {title} {\bibinfo {title} {Lattice dynamics and dathermophysical properties of h.c.p. re and tc from the quasi-harmonic approximation},\ }\href@noop {} {\bibfield  {journal} {\bibinfo  {journal} {Physica Status Solidi (B)}\ }\textbf {\bibinfo {volume} {254}},\ \bibinfo {pages} {1700101} (\bibinfo {year} {2017}{\natexlab{b}})}\BibitemShut {NoStop}%
\bibitem [{\citenamefont {Malica}\ and\ \citenamefont {Dal~Corso}(2019)}]{malica_temperature_dependent_2019}%
  \BibitemOpen
  \bibfield  {author} {\bibinfo {author} {\bibfnamefont {C.}~\bibnamefont {Malica}}\ and\ \bibinfo {author} {\bibfnamefont {A.}~\bibnamefont {Dal~Corso}},\ }\bibfield  {title} {\bibinfo {title} {Temperature-dependent atomic {B} factor: an ab initio calculation},\ }\href {https://doi.org/10.1107/S205327331900514X} {\bibfield  {journal} {\bibinfo  {journal} {Acta Crystallographica Section A}\ }\textbf {\bibinfo {volume} {75}},\ \bibinfo {pages} {624} (\bibinfo {year} {2019})}\BibitemShut {NoStop}%
\bibitem [{\citenamefont {Malica}\ and\ \citenamefont {Dal~Corso}(2020{\natexlab{b}})}]{malica_temperature_2020}%
  \BibitemOpen
  \bibfield  {author} {\bibinfo {author} {\bibfnamefont {C.}~\bibnamefont {Malica}}\ and\ \bibinfo {author} {\bibfnamefont {A.}~\bibnamefont {Dal~Corso}},\ }\bibfield  {title} {\bibinfo {title} {Temperature dependent elastic constants and thermodynamic properties of {BAs}: {An} ab initio investigation},\ }\href@noop {} {\bibfield  {journal} {\bibinfo  {journal} {Journal of Applied Physics}\ }\textbf {\bibinfo {volume} {127}},\ \bibinfo {pages} {245103} (\bibinfo {year} {2020}{\natexlab{b}})}\BibitemShut {NoStop}%
\bibitem [{\citenamefont {Barron}\ and\ \citenamefont {Klein}(1965)}]{barron_second-order_1965}%
  \BibitemOpen
  \bibfield  {author} {\bibinfo {author} {\bibfnamefont {T.~H.~K.}\ \bibnamefont {Barron}}\ and\ \bibinfo {author} {\bibfnamefont {M.~L.}\ \bibnamefont {Klein}},\ }\bibfield  {title} {\bibinfo {title} {Second-order elastic constants of a solid under stress},\ }\href {https://doi.org/10.1088/0370-1328/85/3/313} {\bibfield  {journal} {\bibinfo  {journal} {Proceedings of the Physical Society}\ }\textbf {\bibinfo {volume} {85}},\ \bibinfo {pages} {523} (\bibinfo {year} {1965})}\BibitemShut {NoStop}%
\bibitem [{\citenamefont {Malica}\ and\ \citenamefont {Dal~Corso}(2022)}]{malicaFinitetemperatureAtomicRelaxations2022}%
  \BibitemOpen
  \bibfield  {author} {\bibinfo {author} {\bibfnamefont {C.}~\bibnamefont {Malica}}\ and\ \bibinfo {author} {\bibfnamefont {A.}~\bibnamefont {Dal~Corso}},\ }\bibfield  {title} {\bibinfo {title} {Finite-temperature atomic relaxations: {{Effect}} on the temperature-dependent {{C44}} elastic constants of {{Si}} and {{BAs}}},\ }\href {https://doi.org/10.1063/5.0093376} {\bibfield  {journal} {\bibinfo  {journal} {The Journal of Chemical Physics}\ }\textbf {\bibinfo {volume} {156}},\ \bibinfo {pages} {194111} (\bibinfo {year} {2022})}\BibitemShut {NoStop}%
\bibitem [{\citenamefont {Liu}\ and\ \citenamefont {Allen}(2018)}]{liuInternalExternalThermal2018}%
  \BibitemOpen
  \bibfield  {author} {\bibinfo {author} {\bibfnamefont {J.}~\bibnamefont {Liu}}\ and\ \bibinfo {author} {\bibfnamefont {P.~B.}\ \bibnamefont {Allen}},\ }\bibfield  {title} {\bibinfo {title} {Internal and external thermal expansions of wurtzite {{ZnO}} from first principles},\ }\href {https://doi.org/10.1016/j.commatsci.2018.07.053} {\bibfield  {journal} {\bibinfo  {journal} {Computational Materials Science}\ }\textbf {\bibinfo {volume} {154}},\ \bibinfo {pages} {251} (\bibinfo {year} {2018})}\BibitemShut {NoStop}%
\bibitem [{\citenamefont {Giannozzi}\ \emph {et~al.}(2009)\citenamefont {Giannozzi}, \citenamefont {Baroni}, \citenamefont {Bonini}, \citenamefont {Calandra}, \citenamefont {Car}, \citenamefont {Cavazzoni}, \citenamefont {Ceresoli}, \citenamefont {Chiarotti}, \citenamefont {Cococcioni}, \citenamefont {Dabo}, \citenamefont {Dal~Corso}, \citenamefont {de~Gironcoli}, \citenamefont {Fabris}, \citenamefont {Fratesi}, \citenamefont {Gebauer}, \citenamefont {Gerstmann}, \citenamefont {Gougoussis}, \citenamefont {Kokalj}, \citenamefont {Lazzeri}, \citenamefont {Martin-Samos}, \citenamefont {Marzari}, \citenamefont {Mauri}, \citenamefont {Mazzarello}, \citenamefont {Paolini}, \citenamefont {Pasquarello}, \citenamefont {Paulatto}, \citenamefont {Sbraccia}, \citenamefont {Scandolo}, \citenamefont {Sclauzero}, \citenamefont {Seitsonen}, \citenamefont {Smogunov}, \citenamefont {Umari},\ and\ \citenamefont {Wentzcovitch}}]{qe1}%
  \BibitemOpen
  \bibfield  {author} {\bibinfo {author} {\bibfnamefont {P.}~\bibnamefont {Giannozzi}}, \bibinfo {author} {\bibfnamefont {S.}~\bibnamefont {Baroni}}, \bibinfo {author} {\bibfnamefont {N.}~\bibnamefont {Bonini}}, \bibinfo {author} {\bibfnamefont {M.}~\bibnamefont {Calandra}}, \bibinfo {author} {\bibfnamefont {R.}~\bibnamefont {Car}}, \bibinfo {author} {\bibfnamefont {C.}~\bibnamefont {Cavazzoni}}, \bibinfo {author} {\bibfnamefont {D.}~\bibnamefont {Ceresoli}}, \bibinfo {author} {\bibfnamefont {G.~L.}\ \bibnamefont {Chiarotti}}, \bibinfo {author} {\bibfnamefont {M.}~\bibnamefont {Cococcioni}}, \bibinfo {author} {\bibfnamefont {I.}~\bibnamefont {Dabo}}, \bibinfo {author} {\bibfnamefont {A.}~\bibnamefont {Dal~Corso}}, \bibinfo {author} {\bibfnamefont {S.}~\bibnamefont {de~Gironcoli}}, \bibinfo {author} {\bibfnamefont {S.}~\bibnamefont {Fabris}}, \bibinfo {author} {\bibfnamefont {G.}~\bibnamefont {Fratesi}}, \bibinfo {author} {\bibfnamefont {R.}~\bibnamefont {Gebauer}}, \bibinfo {author} {\bibfnamefont
  {U.}~\bibnamefont {Gerstmann}}, \bibinfo {author} {\bibfnamefont {C.}~\bibnamefont {Gougoussis}}, \bibinfo {author} {\bibfnamefont {A.}~\bibnamefont {Kokalj}}, \bibinfo {author} {\bibfnamefont {M.}~\bibnamefont {Lazzeri}}, \bibinfo {author} {\bibfnamefont {L.}~\bibnamefont {Martin-Samos}}, \bibinfo {author} {\bibfnamefont {N.}~\bibnamefont {Marzari}}, \bibinfo {author} {\bibfnamefont {F.}~\bibnamefont {Mauri}}, \bibinfo {author} {\bibfnamefont {R.}~\bibnamefont {Mazzarello}}, \bibinfo {author} {\bibfnamefont {S.}~\bibnamefont {Paolini}}, \bibinfo {author} {\bibfnamefont {A.}~\bibnamefont {Pasquarello}}, \bibinfo {author} {\bibfnamefont {L.}~\bibnamefont {Paulatto}}, \bibinfo {author} {\bibfnamefont {C.}~\bibnamefont {Sbraccia}}, \bibinfo {author} {\bibfnamefont {S.}~\bibnamefont {Scandolo}}, \bibinfo {author} {\bibfnamefont {G.}~\bibnamefont {Sclauzero}}, \bibinfo {author} {\bibfnamefont {A.~P.}\ \bibnamefont {Seitsonen}}, \bibinfo {author} {\bibfnamefont {A.}~\bibnamefont {Smogunov}}, \bibinfo {author}
  {\bibfnamefont {P.}~\bibnamefont {Umari}},\ and\ \bibinfo {author} {\bibfnamefont {R.~M.}\ \bibnamefont {Wentzcovitch}},\ }\bibfield  {title} {\bibinfo {title} {{QUANTUM} {ESPRESSO}: a modular and open-source software project for quantum simulations of materials},\ }\href@noop {} {\bibfield  {journal} {\bibinfo  {journal} {Journal of Physics: Condensed Matter}\ }\textbf {\bibinfo {volume} {21}},\ \bibinfo {pages} {395502} (\bibinfo {year} {2009})}\BibitemShut {NoStop}%
\bibitem [{\citenamefont {Giannozzi}\ \emph {et~al.}(2017)\citenamefont {Giannozzi}, \citenamefont {Andreussi}, \citenamefont {Brumme}, \citenamefont {Bunau}, \citenamefont {Buongiorno~Nardelli}, \citenamefont {Calandra}, \citenamefont {Car}, \citenamefont {Cavazzoni}, \citenamefont {Ceresoli}, \citenamefont {Cococcioni}, \citenamefont {Colonna}, \citenamefont {Carnimeo}, \citenamefont {Dal~Corso}, \citenamefont {de~Gironcoli}, \citenamefont {Delugas}, \citenamefont {DiStasio}, \citenamefont {Ferretti}, \citenamefont {Floris}, \citenamefont {Fratesi}, \citenamefont {Fugallo}, \citenamefont {Gebauer}, \citenamefont {Gerstmann}, \citenamefont {Giustino}, \citenamefont {Gorni}, \citenamefont {Jia}, \citenamefont {Kawamura}, \citenamefont {Ko}, \citenamefont {Kokalj}, \citenamefont {Küçükbenli}, \citenamefont {Lazzeri}, \citenamefont {Marsili}, \citenamefont {Marzari}, \citenamefont {Mauri}, \citenamefont {Nguyen}, \citenamefont {Nguyen}, \citenamefont {Otero-de-la Roza}, \citenamefont {Paulatto},
  \citenamefont {Poncé}, \citenamefont {Rocca}, \citenamefont {Sabatini}, \citenamefont {Santra}, \citenamefont {Schlipf}, \citenamefont {Seitsonen}, \citenamefont {Smogunov}, \citenamefont {Timrov}, \citenamefont {Thonhauser}, \citenamefont {Umari}, \citenamefont {Vast}, \citenamefont {Wu},\ and\ \citenamefont {Baroni}}]{qe2}%
  \BibitemOpen
  \bibfield  {author} {\bibinfo {author} {\bibfnamefont {P.}~\bibnamefont {Giannozzi}}, \bibinfo {author} {\bibfnamefont {O.}~\bibnamefont {Andreussi}}, \bibinfo {author} {\bibfnamefont {T.}~\bibnamefont {Brumme}}, \bibinfo {author} {\bibfnamefont {O.}~\bibnamefont {Bunau}}, \bibinfo {author} {\bibfnamefont {M.}~\bibnamefont {Buongiorno~Nardelli}}, \bibinfo {author} {\bibfnamefont {M.}~\bibnamefont {Calandra}}, \bibinfo {author} {\bibfnamefont {R.}~\bibnamefont {Car}}, \bibinfo {author} {\bibfnamefont {C.}~\bibnamefont {Cavazzoni}}, \bibinfo {author} {\bibfnamefont {D.}~\bibnamefont {Ceresoli}}, \bibinfo {author} {\bibfnamefont {M.}~\bibnamefont {Cococcioni}}, \bibinfo {author} {\bibfnamefont {N.}~\bibnamefont {Colonna}}, \bibinfo {author} {\bibfnamefont {I.}~\bibnamefont {Carnimeo}}, \bibinfo {author} {\bibfnamefont {A.}~\bibnamefont {Dal~Corso}}, \bibinfo {author} {\bibfnamefont {S.}~\bibnamefont {de~Gironcoli}}, \bibinfo {author} {\bibfnamefont {P.}~\bibnamefont {Delugas}}, \bibinfo {author} {\bibfnamefont
  {R.~A.}\ \bibnamefont {DiStasio}}, \bibinfo {author} {\bibfnamefont {A.}~\bibnamefont {Ferretti}}, \bibinfo {author} {\bibfnamefont {A.}~\bibnamefont {Floris}}, \bibinfo {author} {\bibfnamefont {G.}~\bibnamefont {Fratesi}}, \bibinfo {author} {\bibfnamefont {G.}~\bibnamefont {Fugallo}}, \bibinfo {author} {\bibfnamefont {R.}~\bibnamefont {Gebauer}}, \bibinfo {author} {\bibfnamefont {U.}~\bibnamefont {Gerstmann}}, \bibinfo {author} {\bibfnamefont {F.}~\bibnamefont {Giustino}}, \bibinfo {author} {\bibfnamefont {T.}~\bibnamefont {Gorni}}, \bibinfo {author} {\bibfnamefont {J.}~\bibnamefont {Jia}}, \bibinfo {author} {\bibfnamefont {M.}~\bibnamefont {Kawamura}}, \bibinfo {author} {\bibfnamefont {H.-Y.}\ \bibnamefont {Ko}}, \bibinfo {author} {\bibfnamefont {A.}~\bibnamefont {Kokalj}}, \bibinfo {author} {\bibfnamefont {E.}~\bibnamefont {Küçükbenli}}, \bibinfo {author} {\bibfnamefont {M.}~\bibnamefont {Lazzeri}}, \bibinfo {author} {\bibfnamefont {M.}~\bibnamefont {Marsili}}, \bibinfo {author} {\bibfnamefont
  {N.}~\bibnamefont {Marzari}}, \bibinfo {author} {\bibfnamefont {F.}~\bibnamefont {Mauri}}, \bibinfo {author} {\bibfnamefont {N.~L.}\ \bibnamefont {Nguyen}}, \bibinfo {author} {\bibfnamefont {H.-V.}\ \bibnamefont {Nguyen}}, \bibinfo {author} {\bibfnamefont {A.}~\bibnamefont {Otero-de-la Roza}}, \bibinfo {author} {\bibfnamefont {L.}~\bibnamefont {Paulatto}}, \bibinfo {author} {\bibfnamefont {S.}~\bibnamefont {Poncé}}, \bibinfo {author} {\bibfnamefont {D.}~\bibnamefont {Rocca}}, \bibinfo {author} {\bibfnamefont {R.}~\bibnamefont {Sabatini}}, \bibinfo {author} {\bibfnamefont {B.}~\bibnamefont {Santra}}, \bibinfo {author} {\bibfnamefont {M.}~\bibnamefont {Schlipf}}, \bibinfo {author} {\bibfnamefont {A.~P.}\ \bibnamefont {Seitsonen}}, \bibinfo {author} {\bibfnamefont {A.}~\bibnamefont {Smogunov}}, \bibinfo {author} {\bibfnamefont {I.}~\bibnamefont {Timrov}}, \bibinfo {author} {\bibfnamefont {T.}~\bibnamefont {Thonhauser}}, \bibinfo {author} {\bibfnamefont {P.}~\bibnamefont {Umari}}, \bibinfo {author}
  {\bibfnamefont {N.}~\bibnamefont {Vast}}, \bibinfo {author} {\bibfnamefont {X.}~\bibnamefont {Wu}},\ and\ \bibinfo {author} {\bibfnamefont {S.}~\bibnamefont {Baroni}},\ }\bibfield  {title} {\bibinfo {title} {Advanced capabilities for materials modelling with {Quantum} {ESPRESSO}},\ }\href {https://www.osti.gov/pages/biblio/1523470} {\bibfield  {journal} {\bibinfo  {journal} {Journal of Physics. Condensed Matter}\ }\textbf {\bibinfo {volume} {29}},\ \bibinfo {pages} {465901} (\bibinfo {year} {2017})}\BibitemShut {NoStop}%
\bibitem [{\citenamefont {Perdew}\ and\ \citenamefont {Zunger}(1981)}]{lda}%
  \BibitemOpen
  \bibfield  {author} {\bibinfo {author} {\bibfnamefont {J.~P.}\ \bibnamefont {Perdew}}\ and\ \bibinfo {author} {\bibfnamefont {A.}~\bibnamefont {Zunger}},\ }\bibfield  {title} {\bibinfo {title} {Self-interaction correction to density-functional approximations for many-electron systems},\ }\href {https://doi.org/10.1103/PhysRevB.23.5048} {\bibfield  {journal} {\bibinfo  {journal} {Physical Review B}\ }\textbf {\bibinfo {volume} {23}},\ \bibinfo {pages} {5048} (\bibinfo {year} {1981})}\BibitemShut {NoStop}%
\bibitem [{\citenamefont {Louie}\ \emph {et~al.}(1982)\citenamefont {Louie}, \citenamefont {Froyen},\ and\ \citenamefont {Cohen}}]{louie_nonlinear_1982}%
  \BibitemOpen
  \bibfield  {author} {\bibinfo {author} {\bibfnamefont {S.~G.}\ \bibnamefont {Louie}}, \bibinfo {author} {\bibfnamefont {S.}~\bibnamefont {Froyen}},\ and\ \bibinfo {author} {\bibfnamefont {M.~L.}\ \bibnamefont {Cohen}},\ }\bibfield  {title} {\bibinfo {title} {Nonlinear ionic pseudopotentials in spin-density-functional calculations},\ }\href {https://doi.org/10.1103/PhysRevB.26.1738} {\bibfield  {journal} {\bibinfo  {journal} {Phys. Rev. B}\ }\textbf {\bibinfo {volume} {26}},\ \bibinfo {pages} {1738} (\bibinfo {year} {1982})}\BibitemShut {NoStop}%
\bibitem [{\citenamefont {Methfessel}\ and\ \citenamefont {Paxton}(1989)}]{mp}%
  \BibitemOpen
  \bibfield  {author} {\bibinfo {author} {\bibfnamefont {M.}~\bibnamefont {Methfessel}}\ and\ \bibinfo {author} {\bibfnamefont {A.~T.}\ \bibnamefont {Paxton}},\ }\bibfield  {title} {\bibinfo {title} {High-precision sampling for brillouin-zone integration in metals},\ }\href {https://doi.org/10.1103/PhysRevB.40.3616} {\bibfield  {journal} {\bibinfo  {journal} {Physical Review B}\ }\textbf {\bibinfo {volume} {40}},\ \bibinfo {pages} {3616} (\bibinfo {year} {1989})}\BibitemShut {NoStop}%
\bibitem [{\citenamefont {Baroni}\ \emph {et~al.}(2001)\citenamefont {Baroni}, \citenamefont {de~Gironcoli}, \citenamefont {Dal~Corso},\ and\ \citenamefont {Giannozzi}}]{rmp}%
  \BibitemOpen
  \bibfield  {author} {\bibinfo {author} {\bibfnamefont {S.}~\bibnamefont {Baroni}}, \bibinfo {author} {\bibfnamefont {S.}~\bibnamefont {de~Gironcoli}}, \bibinfo {author} {\bibfnamefont {A.}~\bibnamefont {Dal~Corso}},\ and\ \bibinfo {author} {\bibfnamefont {P.}~\bibnamefont {Giannozzi}},\ }\bibfield  {title} {\bibinfo {title} {Phonons and related crystal properties from density-functional perturbation theory},\ }\href {https://doi.org/10.1103/RevModPhys.73.515} {\bibfield  {journal} {\bibinfo  {journal} {Review of Modern Physics}\ }\textbf {\bibinfo {volume} {73}},\ \bibinfo {pages} {515} (\bibinfo {year} {2001})}\BibitemShut {NoStop}%
\bibitem [{\citenamefont {Dal~Corso}(2010)}]{dfptPAW}%
  \BibitemOpen
  \bibfield  {author} {\bibinfo {author} {\bibfnamefont {A.}~\bibnamefont {Dal~Corso}},\ }\bibfield  {title} {\bibinfo {title} {Density functional perturbation theory within the projector augmented wave method},\ }\href {https://doi.org/10.1103/PhysRevB.81.075123} {\bibfield  {journal} {\bibinfo  {journal} {Physical Review B}\ }\textbf {\bibinfo {volume} {81}},\ \bibinfo {pages} {075123} (\bibinfo {year} {2010})}\BibitemShut {NoStop}%
\bibitem [{\citenamefont {Gong}\ and\ \citenamefont {Dal~Corso}(2023)}]{gong_dalcorso_opt}%
  \BibitemOpen
  \bibfield  {author} {\bibinfo {author} {\bibfnamefont {X.}~\bibnamefont {Gong}}\ and\ \bibinfo {author} {\bibfnamefont {A.}~\bibnamefont {Dal~Corso}},\ }\bibfield  {title} {\bibinfo {title} {unpublished}} (\bibinfo {year} {2023})\BibitemShut {NoStop}%
\bibitem [{sup()}]{supplemental}%
  \BibitemOpen
  \href@noop {} {}\bibinfo {howpublished} {See Supplemental Material at URL that contains plots of the thermodynamic properties (EOS, $c/a$ as a function of pressure, thermal expansion, isobaric heat capacity, and bulk modulus), a workflow for the calculation of TDECs of hcp solids and a table with the crystal parameters of the 11 geometries studied along the ``stress-pressure'' $T=0$ K isotherm.}\BibitemShut {Stop}%
\bibitem [{\citenamefont {Rostami}\ and\ \citenamefont {Gonze}(2024)}]{rostamiApproximationsFirstprinciplesVolumetric2024}%
  \BibitemOpen
  \bibfield  {author} {\bibinfo {author} {\bibfnamefont {S.}~\bibnamefont {Rostami}}\ and\ \bibinfo {author} {\bibfnamefont {X.}~\bibnamefont {Gonze}},\ }\bibfield  {title} {\bibinfo {title} {Approximations in first-principles volumetric thermal expansion determination},\ }\href {https://doi.org/10.1103/PhysRevB.110.014103} {\bibfield  {journal} {\bibinfo  {journal} {Physical Review B}\ }\textbf {\bibinfo {volume} {110}},\ \bibinfo {pages} {014103} (\bibinfo {year} {2024})}\BibitemShut {NoStop}%
\bibitem [{\citenamefont {Mathis}\ \emph {et~al.}(2022)\citenamefont {Mathis}, \citenamefont {Khanolkar}, \citenamefont {Fu}, \citenamefont {Bryan}, \citenamefont {Dennett}, \citenamefont {Rickert}, \citenamefont {Mann}, \citenamefont {Winn}, \citenamefont {Abernathy}, \citenamefont {Manley}, \citenamefont {Hurley},\ and\ \citenamefont {Marianetti}}]{mathisGeneralizedQuasiharmonicApproximation2022}%
  \BibitemOpen
  \bibfield  {author} {\bibinfo {author} {\bibfnamefont {M.~A.}\ \bibnamefont {Mathis}}, \bibinfo {author} {\bibfnamefont {A.}~\bibnamefont {Khanolkar}}, \bibinfo {author} {\bibfnamefont {L.}~\bibnamefont {Fu}}, \bibinfo {author} {\bibfnamefont {M.~S.}\ \bibnamefont {Bryan}}, \bibinfo {author} {\bibfnamefont {C.~A.}\ \bibnamefont {Dennett}}, \bibinfo {author} {\bibfnamefont {K.}~\bibnamefont {Rickert}}, \bibinfo {author} {\bibfnamefont {J.~M.}\ \bibnamefont {Mann}}, \bibinfo {author} {\bibfnamefont {B.}~\bibnamefont {Winn}}, \bibinfo {author} {\bibfnamefont {D.~L.}\ \bibnamefont {Abernathy}}, \bibinfo {author} {\bibfnamefont {M.~E.}\ \bibnamefont {Manley}}, \bibinfo {author} {\bibfnamefont {D.~H.}\ \bibnamefont {Hurley}},\ and\ \bibinfo {author} {\bibfnamefont {C.~A.}\ \bibnamefont {Marianetti}},\ }\bibfield  {title} {\bibinfo {title} {Generalized quasiharmonic approximation via space group irreducible derivatives},\ }\href {https://doi.org/10.1103/PhysRevB.106.014314} {\bibfield  {journal} {\bibinfo
  {journal} {Physical Review B}\ }\textbf {\bibinfo {volume} {106}},\ \bibinfo {pages} {014314} (\bibinfo {year} {2022})}\BibitemShut {NoStop}%
\bibitem [{\citenamefont {Evans}\ \emph {et~al.}(2005)\citenamefont {Evans}, \citenamefont {Lipp}, \citenamefont {Cynn}, \citenamefont {Yoo}, \citenamefont {Somayazulu}, \citenamefont {H{\"a}usermann}, \citenamefont {Shen},\ and\ \citenamefont {Prakapenka}}]{evansXrayDiffractionRaman2005}%
  \BibitemOpen
  \bibfield  {author} {\bibinfo {author} {\bibfnamefont {W.~J.}\ \bibnamefont {Evans}}, \bibinfo {author} {\bibfnamefont {M.~J.}\ \bibnamefont {Lipp}}, \bibinfo {author} {\bibfnamefont {H.}~\bibnamefont {Cynn}}, \bibinfo {author} {\bibfnamefont {C.~S.}\ \bibnamefont {Yoo}}, \bibinfo {author} {\bibfnamefont {M.}~\bibnamefont {Somayazulu}}, \bibinfo {author} {\bibfnamefont {D.}~\bibnamefont {H{\"a}usermann}}, \bibinfo {author} {\bibfnamefont {G.}~\bibnamefont {Shen}},\ and\ \bibinfo {author} {\bibfnamefont {V.}~\bibnamefont {Prakapenka}},\ }\bibfield  {title} {\bibinfo {title} {X-ray diffraction and {{Raman}} studies of beryllium: {{Static}} and elastic properties at high pressures},\ }\href {https://doi.org/10.1103/PhysRevB.72.094113} {\bibfield  {journal} {\bibinfo  {journal} {Physical Review B}\ }\textbf {\bibinfo {volume} {72}},\ \bibinfo {pages} {094113} (\bibinfo {year} {2005})}\BibitemShut {NoStop}%
\bibitem [{\citenamefont {Mackay}\ and\ \citenamefont {Hill}(1963)}]{mackayLatticeParameterHardness1963}%
  \BibitemOpen
  \bibfield  {author} {\bibinfo {author} {\bibfnamefont {K.~J.~H.}\ \bibnamefont {Mackay}}\ and\ \bibinfo {author} {\bibfnamefont {N.~A.}\ \bibnamefont {Hill}},\ }\bibfield  {title} {\bibinfo {title} {Lattice parameter and hardness measurements on high purity beryllium},\ }\href {https://doi.org/10.1016/0022-3115(63)90043-6} {\bibfield  {journal} {\bibinfo  {journal} {Journal of Nuclear Materials}\ }\textbf {\bibinfo {volume} {8}},\ \bibinfo {pages} {263} (\bibinfo {year} {1963})}\BibitemShut {NoStop}%
\bibitem [{\citenamefont {Perdew}\ \emph {et~al.}(1996)\citenamefont {Perdew}, \citenamefont {Burke},\ and\ \citenamefont {Ernzerhof}}]{pbe}%
  \BibitemOpen
  \bibfield  {author} {\bibinfo {author} {\bibfnamefont {J.~P.}\ \bibnamefont {Perdew}}, \bibinfo {author} {\bibfnamefont {K.}~\bibnamefont {Burke}},\ and\ \bibinfo {author} {\bibfnamefont {M.}~\bibnamefont {Ernzerhof}},\ }\bibfield  {title} {\bibinfo {title} {Generalized gradient approximation made simple},\ }\href {https://doi.org/10.1103/PhysRevLett.77.3865} {\bibfield  {journal} {\bibinfo  {journal} {Physical Review Letters}\ }\textbf {\bibinfo {volume} {77}},\ \bibinfo {pages} {3865} (\bibinfo {year} {1996})}\BibitemShut {NoStop}%
\bibitem [{\citenamefont {Liu}\ \emph {et~al.}(2009)\citenamefont {Liu}, \citenamefont {Liu}, \citenamefont {Whitaker}, \citenamefont {Zhao},\ and\ \citenamefont {Li}}]{liu_experimental_2009}%
  \BibitemOpen
  \bibfield  {author} {\bibinfo {author} {\bibfnamefont {W.}~\bibnamefont {Liu}}, \bibinfo {author} {\bibfnamefont {Q.}~\bibnamefont {Liu}}, \bibinfo {author} {\bibfnamefont {M.~L.}\ \bibnamefont {Whitaker}}, \bibinfo {author} {\bibfnamefont {Y.}~\bibnamefont {Zhao}},\ and\ \bibinfo {author} {\bibfnamefont {B.}~\bibnamefont {Li}},\ }\bibfield  {title} {\bibinfo {title} {Experimental and theoretical studies on the elasticity of molybdenum to 12 {GPa}},\ }\href {https://doi.org/10.1063/1.3197135} {\bibfield  {journal} {\bibinfo  {journal} {Journal of Applied Physics}\ }\textbf {\bibinfo {volume} {106}},\ \bibinfo {pages} {043506} (\bibinfo {year} {2009})}\BibitemShut {NoStop}%
\bibitem [{\citenamefont {Gordon}(1949)}]{gordon_high_1949}%
  \BibitemOpen
  \bibfield  {author} {\bibinfo {author} {\bibfnamefont {P.}~\bibnamefont {Gordon}},\ }\bibfield  {title} {\bibinfo {title} {A high temperature precision x‐ray camera: Some measurements of the thermal coefficients of expansion of beryllium},\ }\href {https://doi.org/10.1063/1.1698252} {\bibfield  {journal} {\bibinfo  {journal} {Journal of Applied Physics}\ }\textbf {\bibinfo {volume} {20}},\ \bibinfo {pages} {908} (\bibinfo {year} {1949})}\BibitemShut {NoStop}%
\bibitem [{\citenamefont {Nakano}\ \emph {et~al.}(2002)\citenamefont {Nakano}, \citenamefont {Akahama},\ and\ \citenamefont {Kawamura}}]{nakanoXrayDiffractionStudy2002}%
  \BibitemOpen
  \bibfield  {author} {\bibinfo {author} {\bibfnamefont {K.}~\bibnamefont {Nakano}}, \bibinfo {author} {\bibfnamefont {Y.}~\bibnamefont {Akahama}},\ and\ \bibinfo {author} {\bibfnamefont {H.}~\bibnamefont {Kawamura}},\ }\bibfield  {title} {\bibinfo {title} {X-ray diffraction study of {{Be}} to megabar pressure},\ }\href {https://doi.org/10.1088/0953-8984/14/44/334} {\bibfield  {journal} {\bibinfo  {journal} {Journal of Physics: Condensed Matter}\ }\textbf {\bibinfo {volume} {14}},\ \bibinfo {pages} {10569} (\bibinfo {year} {2002})}\BibitemShut {NoStop}%
\bibitem [{\citenamefont {Song}\ and\ \citenamefont {Liu}(2007)}]{song_modified_2007}%
  \BibitemOpen
  \bibfield  {author} {\bibinfo {author} {\bibfnamefont {H.-F.}\ \bibnamefont {Song}}\ and\ \bibinfo {author} {\bibfnamefont {H.-F.}\ \bibnamefont {Liu}},\ }\bibfield  {title} {\bibinfo {title} {Modified mean-field potential approach to thermodynamic properties of a low-symmetry crystal: Beryllium as a prototype},\ }\href {https://doi.org/10.1103/PhysRevB.75.245126} {\bibfield  {journal} {\bibinfo  {journal} {Physical Review B}\ }\textbf {\bibinfo {volume} {75}},\ \bibinfo {pages} {245126} (\bibinfo {year} {2007})}\BibitemShut {NoStop}%
\bibitem [{\citenamefont {Lazzeri}\ and\ \citenamefont {{de Gironcoli}}(1998)}]{lazzeriAbinitioDynamicalProperties1998}%
  \BibitemOpen
  \bibfield  {author} {\bibinfo {author} {\bibfnamefont {M.}~\bibnamefont {Lazzeri}}\ and\ \bibinfo {author} {\bibfnamefont {S.}~\bibnamefont {{de Gironcoli}}},\ }\bibfield  {title} {\bibinfo {title} {Ab-initio dynamical properties of the {{Be}}(0001) surface},\ }\href {https://doi.org/10.1016/S0039-6028(97)00993-X} {\bibfield  {journal} {\bibinfo  {journal} {Surface Science}\ }\textbf {\bibinfo {volume} {402--404}},\ \bibinfo {pages} {715} (\bibinfo {year} {1998})}\BibitemShut {NoStop}%
\bibitem [{\citenamefont {Robert}\ and\ \citenamefont {Sollier}(2006)}]{robertEquationStateElastic2006}%
  \BibitemOpen
  \bibfield  {author} {\bibinfo {author} {\bibfnamefont {G.}~\bibnamefont {Robert}}\ and\ \bibinfo {author} {\bibfnamefont {A.}~\bibnamefont {Sollier}},\ }\bibfield  {title} {\bibinfo {title} {Equation of state and elastic properties of beryllium from first principles calculations},\ }\href {https://doi.org/10.1051/jp4:2006134039} {\bibfield  {journal} {\bibinfo  {journal} {Journal de Physique IV (Proceedings)}\ }\textbf {\bibinfo {volume} {134}},\ \bibinfo {pages} {257} (\bibinfo {year} {2006})}\BibitemShut {NoStop}%
\bibitem [{\citenamefont {Allen}(2015)}]{allenAnharmonicPhononQuasiparticle2015}%
  \BibitemOpen
  \bibfield  {author} {\bibinfo {author} {\bibfnamefont {P.~B.}\ \bibnamefont {Allen}},\ }\bibfield  {title} {\bibinfo {title} {Anharmonic phonon quasiparticle theory of zero-point and thermal shifts in insulators: {{Heat}} capacity, bulk modulus, and thermal expansion},\ }\href {https://doi.org/10.1103/PhysRevB.92.064106} {\bibfield  {journal} {\bibinfo  {journal} {Physical Review B}\ }\textbf {\bibinfo {volume} {92}},\ \bibinfo {pages} {064106} (\bibinfo {year} {2015})}\BibitemShut {NoStop}%
\end{thebibliography}%
